\documentclass[twocolumn, twocolappendix]{aastex631}

\usepackage{amsmath}

\graphicspath{{./}}

\shorttitle{LAEs at $z \sim 3.7$ and $4.8$}
\shortauthors{Liu et al.}

\begin{document}

\title{A spectroscopic survey of Ly$\alpha$ emitters and Ly$\alpha$ luminosity function at Redshifts 3.7 and 4.8}
\author[0000-0002-4385-0270]{Weiyang Liu}
\affiliation{Department of Astronomy, School of Physics, Peking University, Beijing 100871, China}
\affiliation{Kavli Institude for Astronomy and Astrophysics, Peking University, Beijing 100871, China}

\author[0000-0003-4176-6486]{Linhua Jiang}
\affiliation{Department of Astronomy, School of Physics, Peking University, Beijing 100871, China}
\affiliation{Kavli Institude for Astronomy and Astrophysics, Peking University, Beijing 100871, China}

\begin{abstract}
We present a spectroscopic survey of Ly$\alpha$ emitters (LAEs) at $z\sim3.7$ and $z\sim4.8$. The LAEs are selected using the narrowband technique based on the combination of deep narrowband and broadband imaging data in two deep fields, and then spectroscopically confirmed with the MMT multi-fiber spectrograph Hectospec. The sample consists of 71 LAEs at $z\sim3.7$ and 69 LAEs at $z\sim4.8$ over $\sim 1.5$ deg$^2$, making it one of the largest spectroscopically confirmed sample of LAEs at the two redshifts. Their Ly$\alpha$ luminosities are measured using the secure redshifts and deep photometric data, and span a range of $\sim 10^{42.5}$ - $10^{43.6} \,\rm erg\, s^{-1}$, so these LAEs represent the most luminous galaxies at the redshifts in terms of Ly$\alpha$ luminosity. We estimate and correct sample incompletenesses and derive reliable Ly$\alpha$ luminosity function (LF)s at $z\sim3.7$ and 4.8 based on the two spectroscopic samples. We find that our Ly$\alpha$ LFs are roughly consistent (within a factor of $2-3$) with previous measurements at similar redshifts that were derived from either photometric samples or spectroscopic samples. By comparing with previous studies in different redshifts, we find that the Ly$\alpha$ LFs decrease mildly from $z\sim3.1$ to $z\sim5.7$, supporting the previous claim of the slow LF evolution between $z\sim2$ and $z\sim6$. At $z>5.7$, the LF declines rapidly towards higher redshift, partly due to the effect of cosmic reionization.
\end{abstract}


\section{Introduction} \label{sec:intro}
Spectroscopically confirmed galaxies at high redshift are important for us to understand galaxy properties and evolution in the distant Universe. In the past two decades, galaxies at $z>2$ have been routinely found using the dropout (or Lyman break) technique \citep{1996ApJ...462L..17S, 2016ARA&A..54..761S, 2023ApJS..265....5H}. This technique uses strong Lyman break in the spectra of star-forming galaxies to select Lyman-break galaxy (LBG) candidates. Follow-up spectroscopic confirmation of these objects are often observationally expensive because of their faint continuum emission. In addition to the Lyman break technique, the narrowband technique that uses strong Ly$\alpha$ lines has also played an important role in finding high-redshift galaxies. It was predicted 55 years ago by \cite{1967ApJ...147..868P} that primeval galaxies undergoing their initial burst of intense star formation would appear very bright in the redshifted Ly$\alpha$ line. This is used by the narrowband technique to select Ly$\alpha$ emitting galaxy (Ly$\alpha$ emitter, or LAE) candidates. These candidates are typically much easier to be spectroscopically identified due to their strong Ly$\alpha$ emission. The two techniques are highly complementary.

\begin{figure*}[t]
\plotone{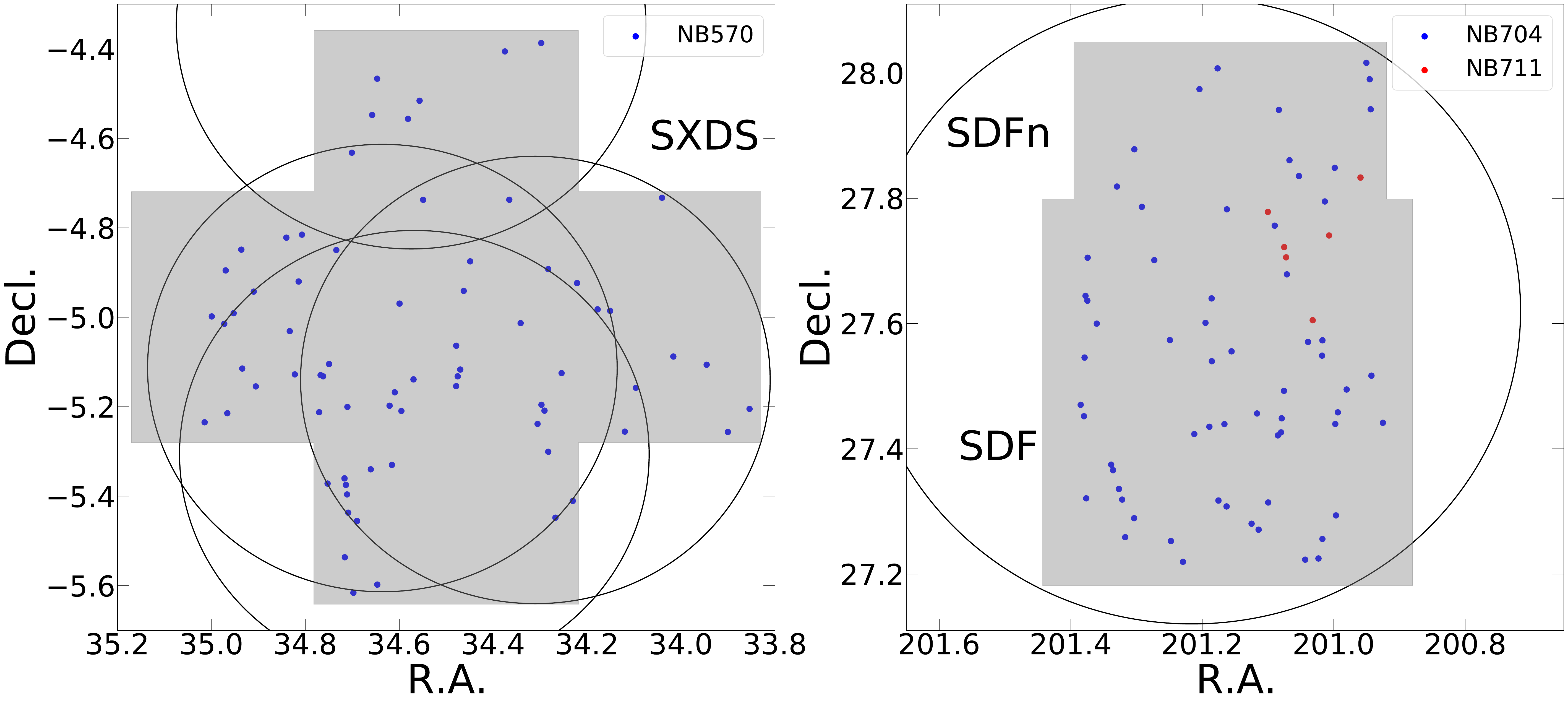}
\caption{LAEs in the SXDS, SDF, and SDFn fields. The left panel shows LAEs at $z\sim3.7$ in SXDS and the right panel shows LAEs at $z\sim4.8$ in SDF and SDFn. In each panel, the grey regions indicate the coverage area of the Subaru Suprime-Cam images. The large circles represent the pointings of our Hectospec observations. The color-coded points represent spectroscopically confirmed LAEs.
\label{fig:fields}}
\end{figure*}

High-redshift LAEs are usually young, compact, metal-poor, and low-mass (stellar mass $\sim 10^{8-9} M_{\odot}$) star-forming galaxies with star formation rates around $\sim 1-10 \, M_{\odot}/ \rm yr$ \citep{2020ARA&A..58..617O}. Since the discovery of the first high-redshift LAEs using the narrowband technique \citep{1996Natur.380..411P, 1996Natur.382..231H, 1996Natur.383...45P}, this technique has successfully found a large number of LAEs at $z\ge2$ \citep[e.g.,][]{2008ApJS..176..301O, 2012ApJ...744..110C, 2018MNRAS.476.4725S, 2019ApJ...886...90H, 2020ApJ...902..137G, 2020ApJ...903....4N, 2022ApJ...926..230N}. Large samples of LAEs from low to high redshift allow us to study their Ly$\alpha$ luminosity function (LF) and evolution \citep[e.g.,][]{2011A&A...525A.143C, 2017A&A...608A...6D, 2018PASJ...70S..16K, 2019A&A...621A.107H, 2020ApJ...902..137G, 2021ApJ...922..167Z}, and physical properties \citep[e.g.,][]{2013ApJ...772...99J, 2016ApJ...816...16J, 2019ApJ...871..164S, 2011ApJ...736..160S}. They also allow us to find protoclusters of galaxies \citep[e.g.,][]{2018NatAs...2..962J, 2021NatAs...5..485H} and characterize the end of cosmic reionization \citep[e.g.,][]{2006ApJ...648....7K, 2011ApJ...734..119K, 2022ApJ...926..230N}. The recently launched James Webb Space Telescope (JWST) allows us to study their rest-frame optical spectral properties and their correlations with the Ly$\alpha$ line \citep[e.g.,][]{2023arXiv230401437R}.

The Ly$\alpha$ LF describes the number density of LAEs as a function of Ly$\alpha$ luminosity, and is thus a basic statistical property of LAEs. Current studies show that the Ly$\alpha$ LF increases rapidly from $z\sim0.3$ to $z\sim3$, appears constant from $z\sim3$ to $z\sim6$, and then declines rapidly from $z\sim6$ towards higher redshift \citep{2020ARA&A..58..617O}. However, the Ly$\alpha$ LF at $z\sim 4-5$ has not been well explored, and previous studies for this redshift range were mostly based on photometric samples of LAEs. 
For example, \cite{2003ApJ...582...60O} calculated $z\sim4.86$ Ly$\alpha$ LF using a photometric sample consisting of 87 LAE candiadtes. \cite{2008ApJS..176..301O} derived $z\sim3.7$ Ly$\alpha$ LF based on 101 photometrically selected LAE candidates, and they spectroscopically confirmed 26 LAEs. \cite{2009ApJ...696..546S} obtained $z\sim4.86$ Ly$\alpha$ LF with 79 photometrically selected LAE candidates. There are some studies based on spectroscopic samples, but the number is relatively small. For example, \cite{2013MNRAS.431.3589Z} combined results from the Large Area Lyman Alpha (LALA) survey and derived $z\sim4.5$ Ly$\alpha$ LF based on a large sample of 207 spectroscopically confirmed LAEs. Although the typical confirmation rate of LAE candidates selected by the narrowband technique is high \citep[$\sim60\%$ to $\sim80\%$, e.g.,][]{2011ApJ...734..119K, 2013MNRAS.431.3589Z}, a spectroscopically confirmed LAE sample is still important in studying Ly$\alpha$ LF by excluding contaminants and deriving more robust Ly$\alpha$ flux. 

In this paper, we present spectroscopic surveys of LAEs at $z\sim3.7$ in the Subaru XMM-Newton Deep Survey (SXDS) field and LAEs at $z\sim4.8$ in the Subaru Deep Field (SDF). We select LAE candidates from deep broadband and narrowband images taken by the Subaru Suprime-Cam and spectroscopically observe them with the MMT Hectospec spectrograph. From these observations, we obtain large samples of LAEs at $z\sim3.7$ and $z\sim4.8$, and we further derive Ly$\alpha$ LFs at the two redshifts. The layout of this paper is as follows. In Section \ref{sec:imgSpecDat}, we introduce target selection and spectroscopic observations. In Section \ref{sec:samples}, we describe LAE samples and properties. Ly$\alpha$ LFs are derived in Section \ref{sec:laeLF}. Section \ref{sec:discussion} and Section \ref{sec:summary} are discussion and summary. Throughout this paper, all magnitudes are in the AB system. We adopt a $\Lambda$-dominated flat cosmology with $H_0=70 \rm \, km \, s^{-1} \, Mpc^{-1}$, $\Omega_m=0.3$, and $\Omega_\Lambda=0.7$. 

\section{Target selection and spectroscopic observations} \label{sec:imgSpecDat}

In this section, we will describe the imaging data in the SXDS and SDF fields, the selection of LAE candidates, the spectroscopic observations of these candidates, and our data reduction.

\subsection{Subaru Suprime-Cam images in SXDS and SDF}

The Suprime-Cam is a wide-field prime-focus imager with a field-of-view of $34\arcmin \times 27\arcmin$ and a pixel scale of $0.202\arcsec$ per pixel for the 8.2m Subaru telescope. The Subaru XMM-Newton Deep Survey \citep[SXDS,][Figure \ref{fig:fields}]{2008ApJS..176....1F} is centered on ($\rm 02^h18^m00^s$, $-05\arcdeg00\arcmin00\arcsec$). It consists of five contiguous subfields SXDS-C, N, S, E, and W (hereafter SXDS1, 2, 3, 4, and 5) corresponding to five Suprime-Cam pointings. The total area coverage is $\sim 1.2$ deg$^2$. The Subaru deep field \citep[SDF,][Figure \ref{fig:fields}]{2004PASJ...56.1011K} is centered on ($\rm 13^h24^m38\fs9$, $+27\arcdeg29\arcmin25\farcs9$) and covers an area of $\sim 800$ arcmin$^2$. Suprime-Cam has taken deep images of the SXDS and SDF fields in a series of broad and narrow bands. These images have been widely used to search for high-redshift LBGs and LAEs  \citep[e.g.,][]{2006ApJ...653..988Y, 2008ApJS..176..301O, 2006ApJ...648....7K, 2011ApJ...734..119K, 2014ApJ...797...16K, 2016ApJ...823...20K}.

\begin{figure}[t]
\epsscale{1.15}
\plotone{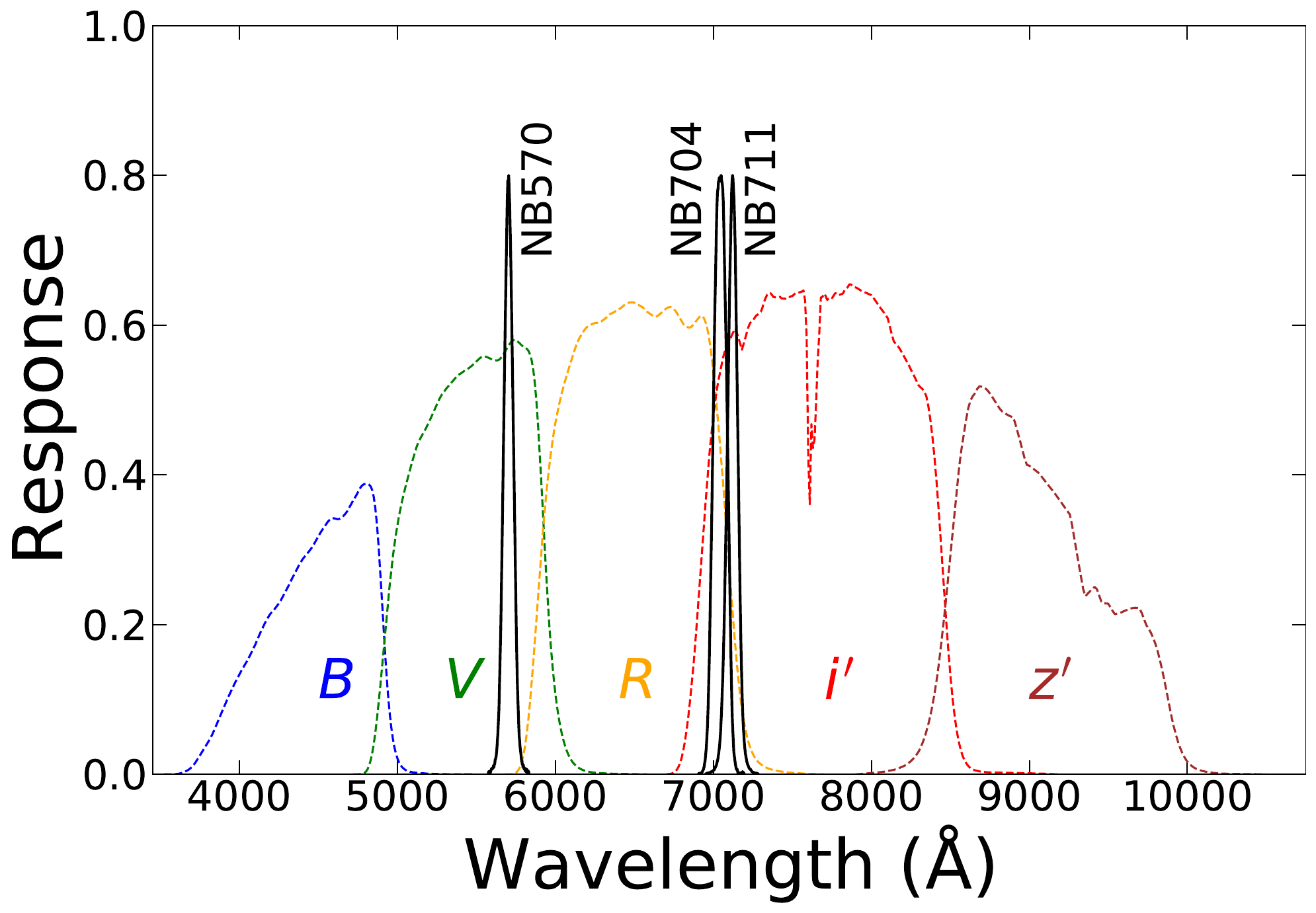}
\caption{Response curves of the narrowband filters (black solid lines) and broadband filters (dashed lines) used in this work. Instrument response has been included for the broad bands.The maximum responses of the narrowband filters are set to 0.8 for the purpose of clarity. The filter NB570 is used to select $z\sim3.7$ LAEs, and the NB704 and NB711 filters are used to select $z\sim4.8$ LAEs. 
\label{fig:filter}}
\end{figure}

We retrieved the raw images from the archival server SMOKA \citep{2002ASPC..281..298B}. The images were reduced, re-sampled, and co-added using a combination of the Suprime-Cam Deep Field REDuction package \citep{2002AJ....123...66Y} and the IDL routines by \citet{2013ApJ...772...99J}. Source extraction and astrometric and photometric solutions were then applied. The details are given in \citet{2013ApJ...772...99J}. We performed aperture photometry with SExtractor \citep{1996A&AS..117..393B} in dual-image mode using the narrowband images as the detection images. Aperture photometry was measured in a $2\arcsec$ diameter aperture and an aperture correction was then determined from a large number of bright but unsaturated point sources and applied to correct for light loss. The depths ($5\sigma$ in a $2\arcsec$ diameter aperture) of the imaging data in five broad bands $BVRi'z'$ reach 27.9, 27.6, 27.4, 27.4, and 26.2 AB mag in SXDS, and 28.0, 27.2, 28.0, 27.8, and 26.8 AB mag in SDF. The typical PSF FWHM of the images is $0.6\arcsec$ in the $i'$ band. Galactic extinction was corrected using the $E(B-V)$ values at the center of SXDS and SDF, respectively.

Narrowband imaging data were reduced in the same method. Figure \ref{fig:filter} shows the transmission curves of the narrow bands used in this paper. No strong OH sky line lies within the passbands of these narrowband filters.  NB570 ($\lambda_c = \rm 5703 \, \AA$; $\rm FWHM = 68 \, \AA$) is used to select $z\sim3.7$ LAE candidates in SXDS. The depth of the NB570 image is about $24.8$ mag and it slightly varies across the five SXDS subfields ($\pm 0.2 \rm \, mag$). NB704 ($\lambda_c = \rm 7042 \, \AA$; $\rm FWHM = 99 \, \AA$) and NB711 ($\lambda_c = \rm 7120 \, \AA$; $\rm FWHM = 73 \, \AA$) are used to select $z\sim4.8$ LAE candidates in SDF. The depths of the NB704 and NB711 images in SDF are $26.2 \rm \, mag$ and $25.5 \rm \, mag$, respectively. A smaller region in the north of SDF (hereafter SDFn) also has $R$, $i'$, NB704, and NB711-band images with depths of 26.9, 26.5, 26.1, and 25.3 mag, respectively, so $z\sim4.8$ LAE candidates are also selected in this field.

\subsection{Target selection in SXDS and SDF}

Based on the filter curves shown in Figure \ref{fig:filter} and the positions of the redshifted Ly$\alpha$ emission lines, we apply the following criteria to select LAE candidates at $z\sim 3.7$ and $z\sim 4.8$. These criteria roughly select LAEs with Ly$\alpha$ rest-frame equivalent width $\rm EW_0 \gtrsim 20 \, \AA$.

For $z\sim3.7$ LAE candidates in SXDS, the Ly$\alpha$ line locates around $\rm 5700 \, \AA$, between the $V$ and $R$ bands. The selection criteria are as follows.

(1) $\rm \sigma(NB570) \le 0.156$ (i.e., $>7\sigma$ detection in NB570) and $\rm NB570>18.0$;

(2) $VR\rm-NB570>0.9$, where $VR$ is an AB magnitude calculated from the $V$ and $R$-band flux: $f_{VR}=0.8 \times f_V + 0.2 \times f_R$;

(3) $\sigma(V) \le 0.362$ (i.e., $>3\sigma$ detection in $V$);

(4) $B-V>0.5$.

The above first and second criteria are the major selection criteria. Criteria 3 and 4 are mainly used to remove contaminants without excluding real LAEs at $z\sim3.7$. For $z\sim 4.8$ LAE candidates in SDF, the Ly$\alpha$ line locates around $\rm 7050 \, \AA$, between the $R$ and $i'$ bands. There are two narrowband filters NB704 and NB711. The selection criteria are as follows.

(1a) For NB704: $\rm \sigma(NB704) \le 0.136$ (i.e., $>8\sigma$ detection in NB704) and $\rm 18.0 < NB704 < 25.55$;

(1b) For NB711: $\rm \sigma(NB711) \le 0.156$ (i.e., $>7\sigma$ detection in NB711) and $\rm NB711>18.0$;

(2) $Ri'-\rm NB>0.9$, where $Ri'$ is an AB magnitude calculated from the $R$ and $i'$-band flux: $f_{Ri'}=0.5 \times f_R + 0.5 \times f_{i'}$;

(3) $\sigma(i') \le 0.362$ (i.e., $>3\sigma$ detection in $i'$);

(4) $\sigma(B) > 0.362$ or $\sigma(V) > 0.362$ or $B-i'>2.0$.

Like for the $z\sim3.7$ LAE candidates, the above first and second criteria are the major selection criteria, and Criteria 3 and 4 are used to remove contaminants without excluding real LAEs at $z\sim4.8$. The selection criteria for $z\sim4.8$ LAE candidates in SDFn are as follows (only $R$, $i'$, NB704, and NB711-band images are available).

(1a) For NB704: $\rm \sigma(NB704) \le 0.136$ (i.e., $>8\sigma$ detection in NB704) and $\rm 20.0 < NB704 < 25.55$;

(1b) For NB711: $\rm \sigma(NB711) \le 0.156$ (i.e., $>7\sigma$ detection in NB711) and $\rm NB711>20.0$;

(2) $Ri'-\rm NB>0.9$, where $Ri'$ is an AB magnitude calculated from the $R$ and $i'$-band flux: $f_{Ri'}=0.5 \times f_R + 0.5 \times f_{i'}$;

(3) $\sigma(r) > 0.362$ or $r-i'>0.1$.

In the above selection procedure, we used slightly different narrowband detection limits ($7\sigma$ or $8\sigma$) for different fields. This is to optimize the target surface densities for follow-up spectroscopy, because different narrowband images have different depths. The selected LAE candidates were visually inspected to exclude spurious detections, such as those near the edges of the narrowband images and those contaminated by nearby bright stars. Finally, we selected 112 $z\sim3.7$ LAE candidates in the NB570 image of SXDS, 123 $z\sim4.8$ LAE candidates in the NB704 images of SDF (87) and SDFn (36), and 28 $z\sim4.8$ LAE candidates in the NB711 images of SDF (16) and SDFn (12).

\begin{deluxetable}{ccccc} 
\tabletypesize{\footnotesize}
\tablewidth{0pt}
\tablenum{1}
\tablecaption{Summary of the MMT Hectospec observations. \label{tab:specObsInfo}}
\tablehead{
\colhead{Date} & \colhead{Grating} & \colhead{Center (R.A., Decl.)} & 
\colhead{Exp. Time} & \colhead{No.}
}
\startdata
 & & $z\sim3.7$ & & \\
\hline
2017.09.28 & 600 & $\rm 02^h18^m33^s, -05\arcdeg06\arcmin48\arcsec$ & 120 min& 64\\
2017.09.28 & 600 & $\rm 02^h18^m33^s, -05\arcdeg06\arcmin48\arcsec$ & 60 min& 45\\
2017.10.01 & 600 & $\rm 02^h18^m33^s, -05\arcdeg06\arcmin48\arcsec$ & 90 min& 64\\
2020.10.11 & 270 & $\rm 02^h17^m14^s, -05\arcdeg08\arcmin24\arcsec$ & 200 min& 22\\
2021.10.03 & 270 & $\rm 02^h18^m16^s, -05\arcdeg18\arcmin21\arcsec$ & 180 min& 9\\
2021.12.08 & 270 & $\rm 02^h18^m18^s, -04\arcdeg20\arcmin50\arcsec$ & 225 min& 14\\
\hline
 & & $z\sim4.8$ & & \\
\hline
2018.05.19 & 600 & $\rm 13^h24^m52^s, +27\arcdeg37\arcmin14\arcsec$ & 80 min& 72\\
2019.04.27 & 600 & $\rm 13^h24^m52^s, +27\arcdeg37\arcmin14\arcsec$ & 240 min& 71\\
\enddata
\end{deluxetable}

\subsection{Spectroscopic observations and data reduction}

We carried out follow-up spectroscopic observations of these LAE candidates from 2017 to 2021, using the optical fiber-fed spectrograph Hectospec on the $\rm 6.5 \, m$ telescope MMT \citep{2005PASP..117.1411F}. Hectospec has a large field-of-view of $1\arcdeg$ in diameter with 300 fibers. The diameter of each fiber is $1\farcs5$ and adjacent fibers can be spaced as closely as $20\arcsec$. The observations are summarized in Table \ref{tab:specObsInfo}. 

We observed the $z\sim3.7$ LAE candidates in SXDS using four Hectospec pointings with two different configurations. This is because many fibers were shared by other projects. In 2017, we used the $\rm 600 \, lines \, mm^{-1}$ grating blazed at $\rm \sim6000 \, \AA$, providing a spectral resolution of $\rm \sim2.1 \, \AA$ and a wavelength coverage from $\rm 4050 \, \AA$ to $\rm 6550 \, \AA$. In 2020 and 2021, we used the $\rm 270 \, lines \, mm^{-1}$ grating blazed at $\rm \sim5000 \, \AA$, providing a spectral resolution of $\rm \sim4.8 \, \AA$ and a wavelength coverage from $\rm 3650 \, \AA$ to $\rm 9200 \, \AA$. For the $z\sim4.8$ LAE candidates in SDF and SDFn, one Hectospec pointing was use to cover most of the sky area. They were observed in 2018 and 2019 with the $\rm 600\,lines\,mm^{-1}$ grating that covered a wavelength coverage from $\rm 6050 \, \AA$ to $\rm 8550 \, \AA$. The seeing of these observations varied from $\sim0.7\arcsec$ to $\sim1.5\arcsec$. For each pointing, about 80 fibers were assigned to blank sky regions for background subtraction.

All $z\sim4.8$ LAE candidates and all but one $z\sim3.7$ LAE candidates were covered by the above Hectospec pointings, and the only one $z\sim3.7$ LAE candidate out of the pointings is excluded in the following analyses. The effective areas of the SXDS sub-fields SXDS1, 2, 3, 4, 5 covered by the pointings are 0.232, 0.235, 0.235, 0.167, and $\rm 0.181 \, deg^2$, respectively. The total effective area in SXDS is $\rm 1.050 \, deg^2$. The effective areas in SDF and SDFn covered by the pointings are 0.279 and $\rm 0.125 \, deg^2$, respectively, with a total area of $\rm 0.404 \, deg^2$. 
Due to the fiber collision, not all candidates were spectroscopically observed. In summary, 99 out of 112 $z\sim3.7$ LAE candidates, 96 out of 123 $z\sim4.8$ LAE candidates in NB704, and 12 out of 28 $z\sim4.8$ LAE candidates in NB711 were spectroscopically observed. 
The total exposure time for each target varies from $\rm 1\,h$ to $\rm 7\,h$. This complexity is partly due to the fact that the fibers were shared by different programs, as mentioned earlier. However, our observing strategy ensures that fainter LAE candidates were assigned with longer exposure time so that their Ly$\alpha$ emission lines can be identified if they are real LAEs.

The Hectospec data were processed and reduced using the HSRED\footnote{http://www.mmto.org/hsred-reduction-pipeline/} reduction pipeline. For each exposure, science and lamp images were bias subtracted, flat-fielded, and cosmic ray rejected. Each fiber was traced, and each one-dimensional (1D) spectrum was extracted. Wavelength solution was then derived. Sky spectrum from each sky fiber was checked and poor sky spectra were rejected. An average “supersky” spectrum was obtained from good sky spectra, which was then scaled according to the strength of skylines in individual object’s spectrum and subtracted from it. The resultant 1D, sky-subtracted, and wavelength-calibrated spectra for different exposures of the same object were then re-sampled to the same wavelength grid and co-added with inverse variance weighting to achieve the final 1D spectrum. 

\section{LAE samples and their properties at $z\sim3.7$ and $z\sim4.8$} \label{sec:samples}

In this section, we construct the spectroscopic samples of LAEs at $z\sim3.7$ and $z\sim4.8$. 
We then derive their spectral properties, including redshift, UV continuum flux, and Ly$\alpha$ line flux and EW.
\begin{figure}[t]
\epsscale{1.15}
\plotone{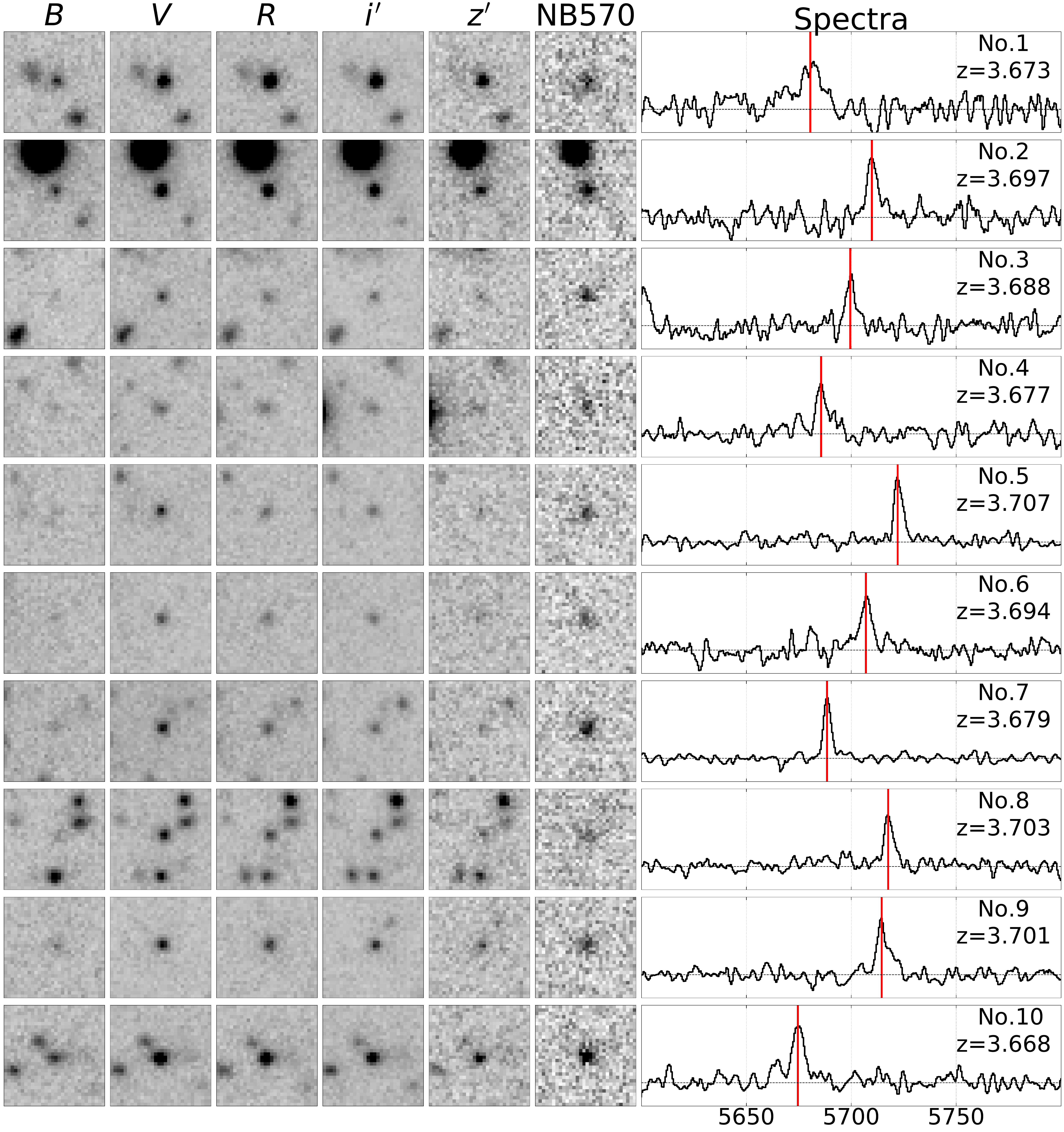}
\caption{Example of 10 LAEs spectroscopically identified at $z\sim3.7$. The left panel shows the stamp images in five broad bands and the narrowband NB570. The image size is $6\arcsec \times 6\arcsec$. The right panel shows the optical spectra of the LAEs taken by MMT Hectospec. The red vertical lines indicate the positions of the redshifted Ly$\alpha$ lines. A figure of the full version is shown in the Appendix.
\label{fig:z37-600}}
\end{figure}

\subsection{LAE samples}

We identify LAEs as follows. For each 1D spectrum, we first search for a possible emission line in the wavelength range covered by the narrowband filters. If a line with signal-to-noise ratio SNR$>5$ is detected, this object 
is regarded as a possible LAE. Low-redshift interlopers are mostly likely [O\,{\scriptsize II}] $\lambda \lambda 3727,3729$, [O\,{\scriptsize III}] $\lambda \lambda 4959,5007$, or H$\beta$ emitters. H$\alpha$ emitters are also possible for $z\sim 4.8$ candidates. These contaminants are identified and excluded using the following steps. If an emission line is [O\,{\scriptsize III}], H$\beta$, or H$\alpha$, the spectrum would cover more than one of these lines, and such an interloper can be easily identified. If an emission line is [O\,{\scriptsize II}] and if the spectrum was taken by the $\rm 270 \, lines \, mm^{-1}$ grating, the spectrum would cover [O\,{\scriptsize III}], H$\beta$, and H$\alpha$, and thus the line is easy to identify. If the spectrum was taken by the $\rm 600 \, lines \, mm^{-1}$ grating, the wavelength coverage is short, but the spectral resolution is higher enough to identify the [O\,{\scriptsize II}] doublet. 
In rare cases, a Ly$\alpha$ line profile can also exhibit a double-peak feature \citep[e.g.,][]{2006A&A...460..397V}, which mimics the [O\,{\scriptsize II}] doublet if its SNR is low. In this case, Ly$\alpha$ and [O\,{\scriptsize II}] can be distinguished using the broadband photometry, given the fact the two redshifts would be very different. If the SNR is high, they can be immediately distinguished by the line profile.
\begin{figure}[t]
\epsscale{1.15}
\plotone{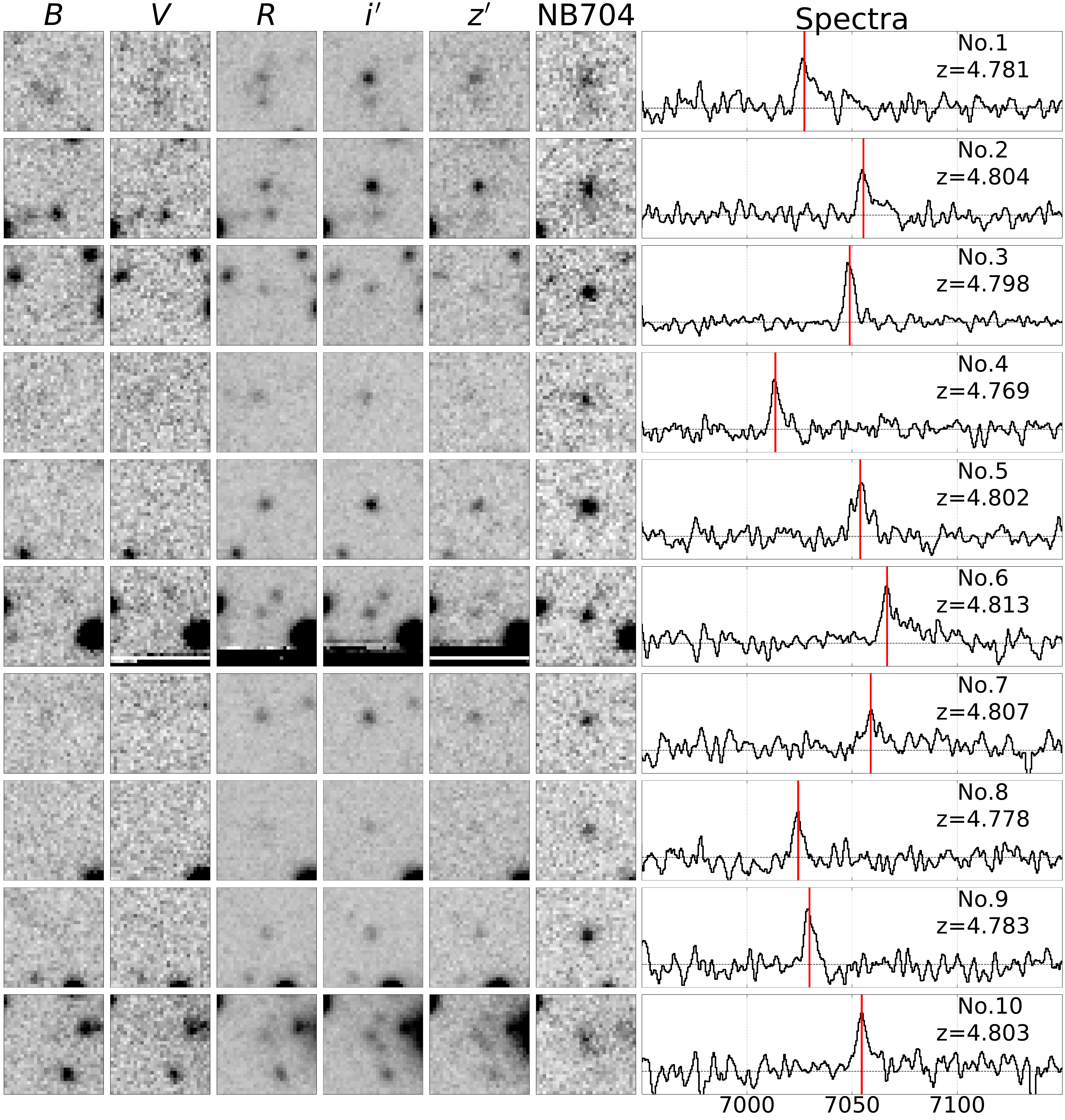}
\caption{Same as Figure \ref{fig:z37-600}, but for our $z\sim4.8$ LAEs selected in NB704. A figure of the full version is shown in the Appendix.
\label{fig:z48-704}}
\end{figure}

Active galactic nuclei (AGNs) are also identified and excluded. For the $z\sim3.7$ LAE candidates in SXDS, deep X-ray images taken by Chandra \citep{2018ApJS..236...48K} and XMM-Newton \citep{2018MNRAS.478.2132C} are available. If an object is detected in X-ray, it is considered as an AGN. In addition, if an object has broad emission lines (line FWHM greater than 1000 km $s^{-1}$) in the spectra, it is also considered as an AGN. One X-ray AGN and one broad-line AGN are identified in our sample.
For the $z\sim4.8$ LAE candidates in SDF and SDFn, there are no X-ray data available, and AGNs are identified based on broad emission lines in the spectra. We find 2 AGNs in our $z\sim4.8$ sample. Thus, the AGN fractions in narrowband selected, LAE photometric samples at both $z\sim3.7$ ($2/99$) and $z\sim4.8$ ($2/(96+12)=2/108$) are around $2\%$, which is consistent with the value reported in previous studies at similar redshifts \citep[e.g.,][]{2008ApJS..176..301O}.

Interestingly, we find that one $z\sim4.8$ LAE candidate ($\rm 13^h25^m32\fs73, +27\arcdeg41\arcmin34\farcs4$) is a supernova (SN) happened in 2001 in a host galaxy at $z=0.515$ (see Figure 7 in \cite{2010PASJ...62...19M}). This object was selected as a LAE candidate because the narrowband image (NB711) was taken in 2001 about one month after the maximum light of the SN, while the broadband images were all taken after 2002 so that the broadband magnitudes were fainter.  

\begin{figure}[t]
\plotone{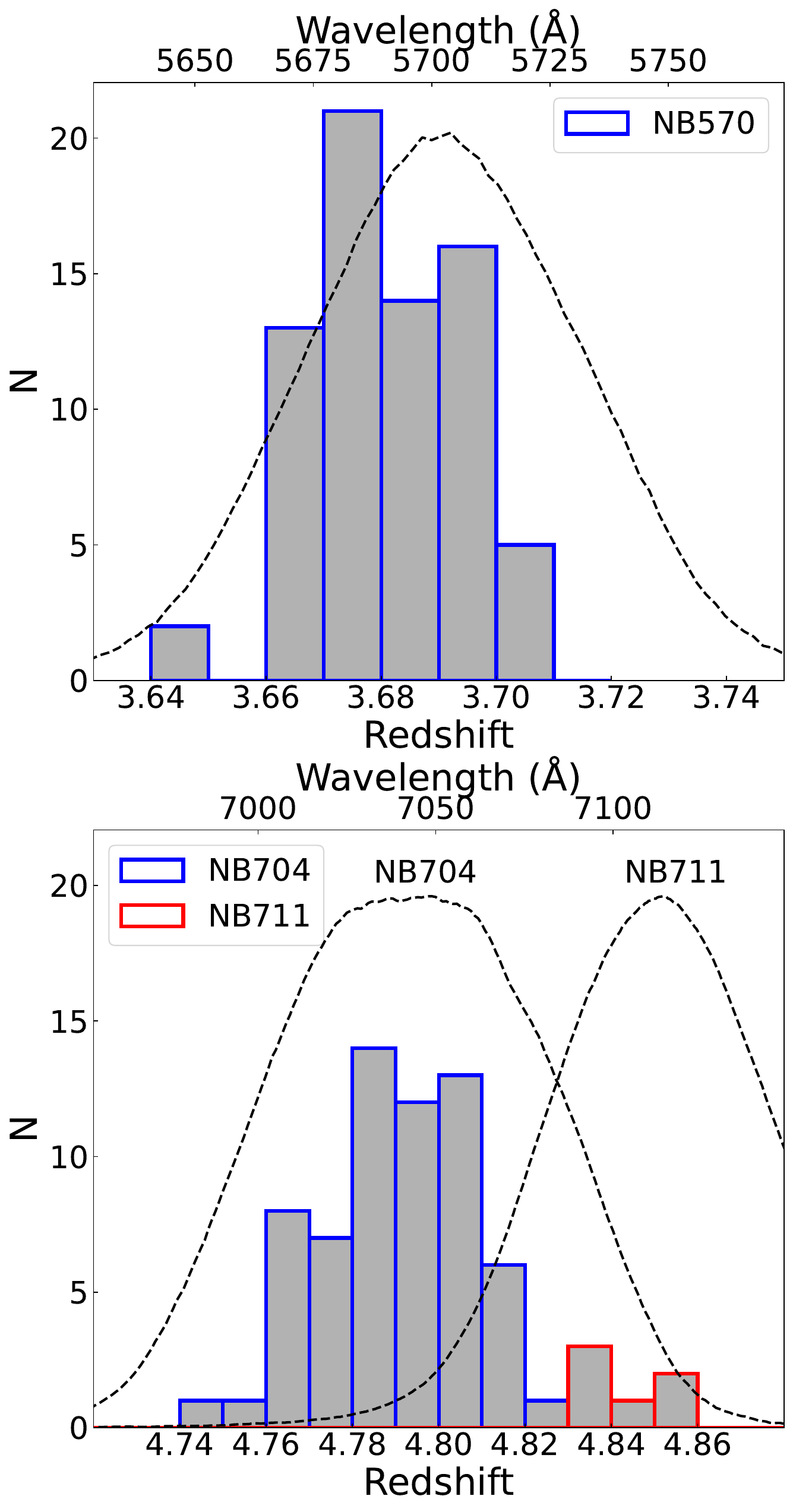}
\caption{Redshift distribution of the spectroscopically confirmed LAEs at $z\sim3.7$ (upper panel) and $z\sim4.8$ (lower panel). The dashed lines show the response curves of the narrowband filters. 
\label{fig:RedshiftDistribution}}
\end{figure}

From the above procedure, 71 $z\sim3.7$ LAEs and 69 $z\sim4.8$ LAEs (including 63 NB704-selected LAEs and 6 NB711-selected LAEs) are spectroscopically confirmed. In the remaining 28 (39) objects at $z\sim3.7$ ($4.8$), we find 1 (3) low-redshift [O\,{\scriptsize II}] emitters, 10 (2) low-redshift [O\,{\scriptsize III}] emitters, and 2 (2) AGNs. In addition, 15 (32) objects do not have SNR$>5$ emission lines in the spectra, and they are likely red stars/galaxies, transients, or spurious detections. In the following analysis, only the confirmed LAEs are considered. Their spatial distributions are shown as the blue and red points in Figure \ref{fig:fields}. Figures \ref{fig:z37-600} presents the stamp images and spectra of the confirmed $z\sim3.7$ LAEs. Figures \ref{fig:z48-704} presents the stamp images and spectra of the confirmed $z\sim4.8$ LAEs. Properties of these LAEs are listed in Table \ref{tab:z37table} and Table \ref{tab:z48table}. 

\begin{deluxetable*}{ccccccccccccc} 
\tabletypesize{\scriptsize}
\tablewidth{0pt}
\tablenum{2}
\tablecaption{Properties of the $z\sim3.7$ LAEs\label{tab:z37table}}
\tablehead{
\colhead{No.} & \colhead{R.A.} & \colhead{Decl.} & \colhead{$B$} & \colhead{$V$} & \colhead{$R$} &
\colhead{$i'$} & \colhead{$z'$} & \colhead{NB} & \colhead{Redshift} &
\colhead{$\log_{10}(L_{\rm Ly\alpha})$} & \colhead{$\rm EW_0$} & \colhead{$\rm M_{UV}$} \\
\colhead{} & \colhead{(J2000)} & \colhead{(J2000)} & \colhead{(mag)} & \colhead{(mag)} & \colhead{(mag)} &
\colhead{(mag)} & \colhead{(mag)} & \colhead{(mag)} & \colhead{} &
\colhead{($\rm erg~s^{-1}$)} & \colhead{($\rm \AA$)} & \colhead{(mag)}
}
\colnumbers
\startdata
1 & 02:18:59.81 & $-$05:06:15.7 & 25.95 & 24.76 & 24.38 & 24.37 & 24.41 & 23.62 & 3.673 & $42.92_{-0.07}^{+0.06}$ & $23.2_{-3.5}^{+3.9}$ & $-21.50_{-0.01}^{+0.01}$\\
2 & 02:18:26.24 & $-$05:10:03.5 & 25.76 & 24.70 & 24.62 & 24.61 & 24.49 & 23.31 & 3.697 & $43.10_{-0.03}^{+0.03}$ & $47.6_{-3.8}^{+4.2}$ & $-21.28_{-0.01}^{+0.01}$\\
3 & 02:18:23.83 & $-$04:58:09.5 & $>$28.46 & 26.29 & 26.59 & 26.60 & $>$26.91 & 23.97 & 3.688 & $42.88_{-0.05}^{+0.05}$ & $135.7_{-38.7}^{+54.4}$ & $-19.27_{-0.10}^{+0.09}$\\
4 & 02:18:16.69 & $-$05:08:19.6 & 27.98 & 26.15 & 26.26 & 26.13 & 25.98 & 24.25 & 3.677 & $42.81_{-0.07}^{+0.07}$ & $117.0_{-25.1}^{+30.6}$ & $-19.68_{-0.06}^{+0.06}$\\
5 & 02:17:54.88 & $-$05:09:14.0 & 27.84 & 25.83 & 26.48 & 26.43 & $>$26.91 & 24.12 & 3.707 & $42.92_{-0.06}^{+0.05}$ & $116.9_{-35.5}^{+48.3}$ & $-19.39_{-0.08}^{+0.08}$\\
6 & 02:17:54.86 & $-$05:03:48.5 & 27.92 & 26.24 & 26.39 & 26.36 & 26.11 & 24.12 & 3.695 & $42.83_{-0.06}^{+0.05}$ & $138.6_{-26.2}^{+32.0}$ & $-19.53_{-0.07}^{+0.07}$\\
7 & 02:17:54.09 & $-$05:07:55.0 & 27.61 & 25.65 & 26.05 & 26.11 & 26.66 & 23.52 & 3.679 & $43.09_{-0.04}^{+0.03}$ & $119.7_{-26.5}^{+33.0}$ & $-19.77_{-0.05}^{+0.05}$\\
8 & 02:17:52.79 & $-$05:07:00.4 & 27.34 & 25.60 & 25.77 & 25.71 & 25.84 & 24.21 & 3.703 & $42.81_{-0.07}^{+0.06}$ & $62.2_{-11.3}^{+13.6}$ & $-20.14_{-0.04}^{+0.04}$\\
9 & 02:17:51.03 & $-$04:56:27.6 & 27.11 & 25.56 & 25.63 & 25.62 & 25.46 & 23.81 & 3.701 & $42.97_{-0.05}^{+0.04}$ & $90.9_{-12.4}^{+14.0}$ & $-20.28_{-0.04}^{+0.04}$\\
10 & 02:17:21.95 & $-$05:00:46.8 & 25.86 & 24.76 & 24.74 & 24.76 & 24.77 & 23.27 & 3.668 & $43.26_{-0.03}^{+0.03}$ & $72.2_{-6.4}^{+7.1}$ & $-21.13_{-0.01}^{+0.01}$\\
11 & 02:17:11.29 & $-$05:11:44.2 & 27.21 & 25.72 & 25.51 & 25.50 & 25.26 & 24.09 & 3.689 & $42.78_{-0.07}^{+0.07}$ & $54.7_{-9.7}^{+11.6}$ & $-20.40_{-0.03}^{+0.03}$\\
12 & 02:17:07.85 & $-$04:53:32.0 & 26.00 & 25.13 & 25.16 & 25.07 & 25.09 & 23.96 & 3.676 & $42.85_{-0.07}^{+0.07}$ & $41.5_{-6.9}^{+7.9}$ & $-20.75_{-0.02}^{+0.02}$\\
13 & 02:17:01.01 & $-$05:07:28.8 & 27.89 & 26.42 & 26.99 & 26.64 & 26.81 & 24.12 & 3.686 & $42.83_{-0.06}^{+0.06}$ & $211.0_{-64.8}^{+92.2}$ & $-19.02_{-0.12}^{+0.12}$\\
14 & 02:18:48.17 & $-$04:37:55.3 & 26.64 & 25.24 & 25.18 & 25.17 & 25.32 & 23.83 & 3.660 & $43.15_{-0.05}^{+0.05}$ & $77.3_{-11.1}^{+12.2}$ & $-20.69_{-0.02}^{+0.02}$\\
15 & 02:18:11.77 & $-$04:44:14.6 & 26.99 & 25.58 & 25.60 & 25.60 & 25.79 & 24.01 & 3.667 & $42.98_{-0.06}^{+0.05}$ & $76.3_{-13.3}^{+15.5}$ & $-20.27_{-0.04}^{+0.04}$\\
16 & 02:17:27.72 & $-$04:44:13.9 & 27.52 & 25.87 & 25.62 & 25.60 & 25.63 & 24.02 & 3.685 & $42.81_{-0.05}^{+0.05}$ & $57.2_{-8.8}^{+9.9}$ & $-20.27_{-0.04}^{+0.04}$\\
17 & 02:19:00.61 & $-$05:22:17.7 & $>$28.76 & 27.53 & 27.75 & 27.35 & $>$26.76 & 24.49 & 3.690 & $42.70_{-0.06}^{+0.06}$ & $429.2_{-181.7}^{+308.7}$ & $-18.33_{-0.25}^{+0.25}$\\
18 & 02:18:51.95 & $-$05:21:36.1 & $>$28.76 & 26.78 & 27.21 & 27.37 & $>$26.76 & 24.47 & 3.672 & $42.81_{-0.06}^{+0.06}$ & $306.9_{-111.7}^{+171.6}$ & $-18.63_{-0.18}^{+0.18}$\\
19 & 02:18:51.79 & $-$05:32:10.9 & 26.15 & 24.73 & 24.47 & 24.44 & 24.34 & 23.42 & 3.671 & $43.11_{-0.03}^{+0.03}$ & $42.7_{-3.6}^{+3.8}$ & $-21.42_{-0.01}^{+0.01}$\\
20 & 02:18:51.24 & $-$05:22:28.5 & 26.53 & 24.98 & 24.65 & 24.50 & 24.55 & 23.80 & 3.672 & $42.89_{-0.06}^{+0.05}$ & $28.8_{-4.2}^{+4.4}$ & $-21.28_{-0.02}^{+0.02}$\\
\enddata
\tablecomments{Col.(1): LAE numbers. Cols.(2)-(3): R.A. and Decl. Cols.(4)-(9): magnitudes in broadband $B$, $V$, $R$, $i'$, $z'$ and in narrowband NB570; $2\sigma$ upper limits are shown if fainter than a $2\sigma$ detection in the filter. Col.(10): redshifts measured from the Ly$\alpha$ line. Col.(11): logarithm of Ly$\alpha$ luminosity in units of $\rm erg~s^{-1}$. Col.(12): rest-frame equivalent width of the Ly$\alpha$ emission line. Col.(13): UV magnitudes at rest-frame 1500$\rm \AA$. Objects without $L_{\rm Ly\alpha}$, $\rm EW_0$, and $\rm M_{UV}$ are affected by nearby objects in their narrow- and broadband photometry. The table only shows the first 20 $z\sim3.7$ LAEs. A full table is available in the electronic version.}
\end{deluxetable*}

\subsection{Ly$\alpha$ redshifts}
We calculate the redshifts of the LAEs using the Ly$\alpha$ emission lines. For each LAE, its redshift is determined by fitting a composite Ly$\alpha$ line profile to the Ly$\alpha$ line in its spectrum. The composite Ly$\alpha$ line profile is obtained as follows. We first assume that the peak of the Ly$\alpha$ line is at $\rm 1215.67 \, \AA$ in the rest frame, and transform all spectra into the rest frame. We then take the average of all spectra to construct a composite Ly$\alpha$ line profile. 
When we fit the composite line profile to the individual Ly$\alpha$ lines, we vary its redshift and amplitude.
After obtaining new redshifts from the fitting results, we use them to transform the spectra into the rest frame again. We repeat the above procedure a few times. The final products include the average Ly$\alpha$ line profiles at $z\sim3.7$ and $z\sim4.8$ and redshifts for all our LAEs. Because the Ly$\alpha$ line is typically redshifted compared with other strong emission lines \citep[e.g.,][]{2003ApJ...588...65S, 2021MNRAS.505.1382M}, possibly due to the back scattering of Ly$\alpha$ photons in outflowing gas \citep[e.g.,][]{2006A&A...460..397V}, Ly$\alpha$ redshifts derived from the Ly$\alpha$ line are higher than systemic redshifts.

\begin{deluxetable*}{ccccccccccccc} 
\tabletypesize{\scriptsize}
\tablewidth{0pt}
\tablenum{3}
\tablecaption{Properties of the $z\sim4.8$ LAEs \label{tab:z48table}}
\tablehead{
\colhead{No.} & \colhead{R.A.} & \colhead{Decl.} & \colhead{$B$} & \colhead{$V$} & \colhead{$R$} &
\colhead{$i'$} & \colhead{$z'$} & \colhead{NB} & \colhead{Redshift} &
\colhead{$\log_{10}(L_{\rm Ly\alpha})$} & \colhead{$\rm EW_0$} & \colhead{$\rm M_{UV}$} \\
\colhead{} & \colhead{(J2000)} & \colhead{(J2000)} & \colhead{(mag)} & \colhead{(mag)} & \colhead{(mag)} &
\colhead{(mag)} & \colhead{(mag)} & \colhead{(mag)} & \colhead{} &
\colhead{($\rm erg~s^{-1}$)} & \colhead{($\rm \AA$)} & \colhead{(mag)}
}
\colnumbers
\startdata
1 & 13:25:32.38 & $+$27:28:13.0 & 28.35 & 27.20 & 26.49 & 25.76 & 25.65 & 25.08 & 4.781 & $42.48_{-0.07}^{+0.07}$ & $20.7_{-3.7}^{+4.0}$ & $-20.52_{-0.05}^{+0.05}$\\
2 & 13:25:31.17 & $+$27:27:07.3 & $>$28.76 & $>$28.20 & 26.46 & 25.53 & 25.47 & 24.62 & 4.804 & $42.78_{-0.04}^{+0.04}$ & $34.5_{-3.3}^{+3.7}$ & $-20.71_{-0.04}^{+0.04}$\\
3 & 13:25:30.95 & $+$27:32:44.8 & $>$28.76 & $>$28.20 & 27.72 & 27.27 & 27.52 & 24.78 & 4.798 & $42.80_{-0.03}^{+0.03}$ & $441.0_{-158.9}^{+326.2}$ & $-17.99_{-0.43}^{+0.58}$\\
4 & 13:25:30.61 & $+$27:38:39.2 & $>$28.76 & $>$28.20 & 28.03 & 27.24 & 27.64 & 25.07 & 4.769 & $42.72_{-0.05}^{+0.05}$ & $253.4_{-81.0}^{+154.6}$ & $-18.39_{-0.34}^{+0.45}$\\
5 & 13:25:30.36 & $+$27:19:15.5 & $>$28.76 & $>$28.20 & 26.74 & 25.90 & 26.07 & 24.18 & 4.802 & $43.02_{-0.02}^{+0.02}$ & $106.8_{-10.2}^{+11.5}$ & $-20.09_{-0.08}^{+0.08}$\\
6 & 13:25:29.95 & $+$27:38:11.7 & $>$28.76 & $>$28.20 & 26.83 & 26.08 & 26.11 & 25.08 & 4.813 & $42.68_{-0.05}^{+0.05}$ & $46.7_{-6.9}^{+8.3}$ & $-20.12_{-0.09}^{+0.09}$\\
7 & 13:25:29.83 & $+$27:42:18.6 & $>$28.76 & $>$28.20 & 27.07 & 26.29 & 26.48 & 25.39 & 4.807 & $42.49_{-0.06}^{+0.06}$ & $35.5_{-5.8}^{+7.2}$ & $-19.95_{-0.07}^{+0.08}$\\
8 & 13:25:26.49 & $+$27:35:59.7 & $>$28.76 & $>$28.20 & 27.92 & 27.44 & $>$27.78 & 25.49 & 4.778 & $42.50_{-0.06}^{+0.06}$ & $144.6_{-40.6}^{+64.5}$ & $-18.44_{-0.27}^{+0.32}$\\
9 & 13:25:21.23 & $+$27:22:29.0 & $>$28.76 & $>$28.20 & 27.47 & 26.95 & 26.86 & 24.96 & 4.783 & $42.71_{-0.04}^{+0.03}$ & $153.6_{-29.9}^{+39.2}$ & $-18.90_{-0.19}^{+0.20}$\\
10 & 13:25:20.55 & $+$27:21:57.1 & $>$28.76 & $>$28.20 & 26.51 & 25.78 & 25.55 & 24.97 & 4.803 & $42.63_{-0.04}^{+0.04}$ & $30.0_{-3.3}^{+3.7}$ & $-20.48_{-0.05}^{+0.05}$\\
11 & 13:25:18.41 & $+$27:20:09.8 & $>$28.76 & $>$28.20 & 27.40 & 26.42 & 26.95 & 24.84 & 4.799 & $42.74_{-0.03}^{+0.03}$ & $84.7_{-10.4}^{+13.1}$ & $-19.64_{-0.10}^{+0.11}$\\
12 & 13:25:17.24 & $+$27:19:08.5 & $>$28.76 & 28.11 & 26.65 & 25.80 & 25.81 & 24.46 & 4.787 & $42.86_{-0.02}^{+0.02}$ & $57.1_{-4.8}^{+5.1}$ & $-20.35_{-0.05}^{+0.05}$\\
13 & 13:25:16.12 & $+$27:15:32.3 & 28.60 & $>$28.20 & 27.55 & 27.47 & $>$27.78 & 25.45 & 4.780 & $42.51_{-0.06}^{+0.06}$ & $160.5_{-49.7}^{+84.5}$ & $-18.37_{-0.31}^{+0.37}$\\
14 & 13:25:12.86 & $+$27:17:21.3 & $>$28.76 & $>$28.20 & 27.17 & 26.48 & $>$27.78 & 24.78 & 4.798 & $42.77_{-0.03}^{+0.03}$ & $100.0_{-13.8}^{+17.0}$ & $-19.53_{-0.12}^{+0.12}$\\
15 & 13:25:05.49 & $+$27:42:04.7 & 27.35 & 27.95 & 26.67 & 26.16 & 26.04 & 25.40 & 4.778 & ... & ... & ...\\
16 & 13:24:59.81 & $+$27:34:24.9 & $>$28.76 & $>$28.20 & 26.97 & 25.87 & 26.17 & 25.00 & 4.812 & $42.69_{-0.04}^{+0.04}$ & $38.1_{-4.1}^{+4.5}$ & $-20.36_{-0.05}^{+0.05}$\\
17 & 13:24:59.44 & $+$27:15:10.3 & $>$28.76 & $>$28.20 & 27.18 & 26.41 & 26.35 & 24.69 & 4.773 & $42.82_{-0.03}^{+0.03}$ & $107.1_{-15.7}^{+18.8}$ & $-19.58_{-0.12}^{+0.13}$\\
18 & 13:24:55.03 & $+$27:13:11.0 & $>$28.76 & $>$28.20 & 26.95 & 26.24 & 25.89 & 24.73 & 4.778 & ... & ... & ...\\
19 & 13:24:50.89 & $+$27:25:25.2 & $>$28.76 & $>$28.20 & 27.59 & 26.87 & 27.23 & 25.12 & 4.793 & $42.63_{-0.04}^{+0.04}$ & $105.9_{-18.0}^{+22.0}$ & $-19.12_{-0.14}^{+0.14}$\\
20 & 13:24:46.81 & $+$27:36:04.2 & $>$28.76 & $>$28.20 & 27.63 & 26.97 & 27.07 & 25.31 & 4.802 & $42.57_{-0.04}^{+0.04}$ & $96.9_{-17.3}^{+22.3}$ & $-19.05_{-0.15}^{+0.16}$\\
\enddata
\tablecomments{Col.(1): LAE numbers. Cols.(2)-(3): R.A. and Decl. Cols.(4)-(9): magnitudes in broadband $B$, $V$, $R$, $i'$, $z'$ and in narrowband NB704 or NB711; $2\sigma$ upper limits are shown if fainter than a $2\sigma$ detection in the filter. Col.(10): redshifts measured from the Ly$\alpha$ line. Col.(11): logarithm of Ly$\alpha$ luminosity in units of $\rm erg~s^{-1}$. Col.(12): rest-frame equivalent width of the Ly$\alpha$ emission line. Col.(13): UV magnitudes at rest-frame 1500$\rm \AA$. Objects without $L_{\rm Ly\alpha}$, $\rm EW_0$, and $\rm M_{UV}$ are affected by nearby objects in their narrow- and broadband photometry. The table only shows the properties of the first 20 $z\sim4.8$ LAEs. A full table is available in the electronic version.}
\end{deluxetable*}

Figure \ref{fig:RedshiftDistribution} shows the redshift distributions of the LAEs at $z\sim3.7$ and $z\sim4.8$. We can see from the figure that, although the narrowband filter response curves are symmetric, the redshift distributions are not. There are more LAEs at lower redshifts. The reason is that the UV continuum at the blue side of Ly$\alpha$ line is much weaker than that at the red side due to the absorption of $\lambda_{\rm rest} < 1215.67\rm \, \AA$ photons by ISM and IGM. Therefore, a relatively lower-redshift LAE tends to be brighter in the narrowband and has a higher possibility to be selected as an LAE candidate.

\subsection{Ly$\alpha$ line and UV continuum flux}

Because the rest-frame UV continuum of these LAEs is too weak to be detected in our spectra, we calculate the Ly$\alpha$ line flux and UV continuum flux based on the redshifts and deep broadband and narrowband photometry \citep{2013ApJ...772...99J}. The UV continuum of star-forming galaxies can typically be represented by a power law \citep[e.g.,][]{1995AJ....110.2665M, 2009ApJ...705..936B, 2014ApJ...793..115B}. In addition, there is no strong emission line other than Ly$\alpha$ in the considered wavelength range \citep[e.g., see a composite spectrum in][]{2003ApJ...588...65S}. Therefore, we can model the UV spectrum of an LAE using
\begin{equation}
f_\lambda = A \times S_{\rm Ly\alpha} + B \times \lambda^{\beta} , \label{eq:specModel}
\end{equation}
where $f_\lambda$ (in unit of $\rm erg \, s^{-1} \, cm^{-2} \, \AA^{-1}$) is the sum of the Ly$\alpha$ emission line flux ($A \times S_{\rm Ly\alpha}$) and the power-law continuum ($B \times \lambda^{\beta}$). Here $S_{\rm Ly\alpha}$ is the average Ly$\alpha$ line profile from Section 3.2, $\beta$ is the UV continuum slope, and $A$ and $B$ are two scaling factors in unit of $\rm erg \, s^{-1} \, cm^{-2} \, \AA^{-1}$. The Ly$\alpha$ profile has negligible impact on the flux estimation because the broad and narrow bands are much wider than typical Ly$\alpha$ line widths. For the wavelength range blueward of Ly$\alpha$, we apply an average IGM absorption to the UV continuum according to \cite{1995ApJ...441...18M} (the composite Ly$\alpha$ line has already taken  into account the IGM absorption). We then fit the model spectra to the observed magnitudes and obtain $A$, $B$, and $\beta$.

\begin{figure}[t]
\epsscale{0.92}
\plotone{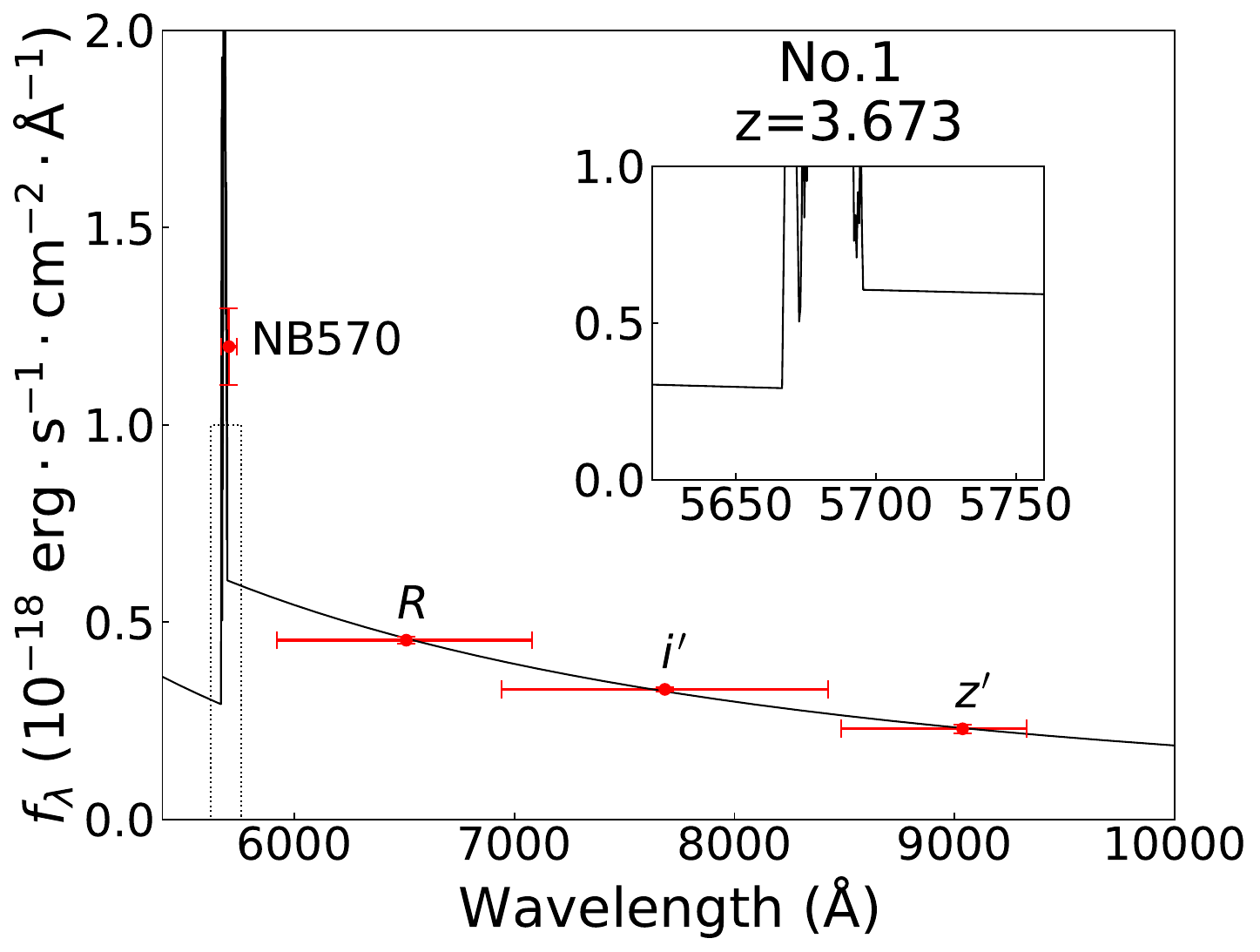}
\caption{Example to illustrate our procedure to model an LAE spectrum and measure its Ly$\alpha$ and UV continuum properties. The red circles are photometric data points. The vertical error bars indicate the photometry errors and the horizontal bars indicate the FWHM of the narrow- and broadband filters. This LAE is at $z=3.673$. Its power-law continuum is fitted with the broadband $R$, $i'$, and $z'$ photometry. After applying the IGM absorption, we calculate the scaling factor $A$ in equation (\ref{eq:specModel}) by matching the Ly$\alpha$ line profile with the narrowband photometry. The inset shows the region around the Ly$\alpha$ line.
\label{fig:z37SpecModel}}
\end{figure}

For the $z\sim3.7$ LAEs, the $R$, $i'$, and $z'$ bands do not cover their Ly$\alpha$ emission lines. Because $f_\lambda \propto \lambda^\beta$ means AB magnitude $m_{\rm AB} \propto (\beta+2) \times \log(\lambda)$, we first use the $R$, $i'$, and $z'$ magnitudes to do a linear fit and obtain $B$ and $\beta$. In rare cases that an object is not detected in a band, we use a $2\sigma$ upper limit for this band in the fitting procedure. We then apply the IGM absorption to the model spectrum and convolve the spectrum with the narrowband transmission curve. Finally, we calculate $A$ by matching the narrowband magnitude from the model spectrum and the observed magnitude. In Figure \ref{fig:z37SpecModel} we present a model fit to a $z\sim3.7$ LAE and illustrate the procedure.

For the $z\sim4.8$ LAEs in SDF and SDFn, the $z'$-band is the only available band redward of Ly$\alpha$ that does not cover the Ly$\alpha$ line, so we are not able to directly calculate continuum slopes. Therefore, we adopt an average UV slope $\beta=-2.17$ according to \cite{2017A&A...608A..10H} for the $z\sim4.8$ LAEs. We calculate $A$ and $B$ using the $i'$ and narrowband magnitudes by comparing the observed magnitudes and the magnitudes measured from the model spectrum. A $2\sigma$ upper limit is used for non-detections, as we did above.

\begin{figure}[t]
\plotone{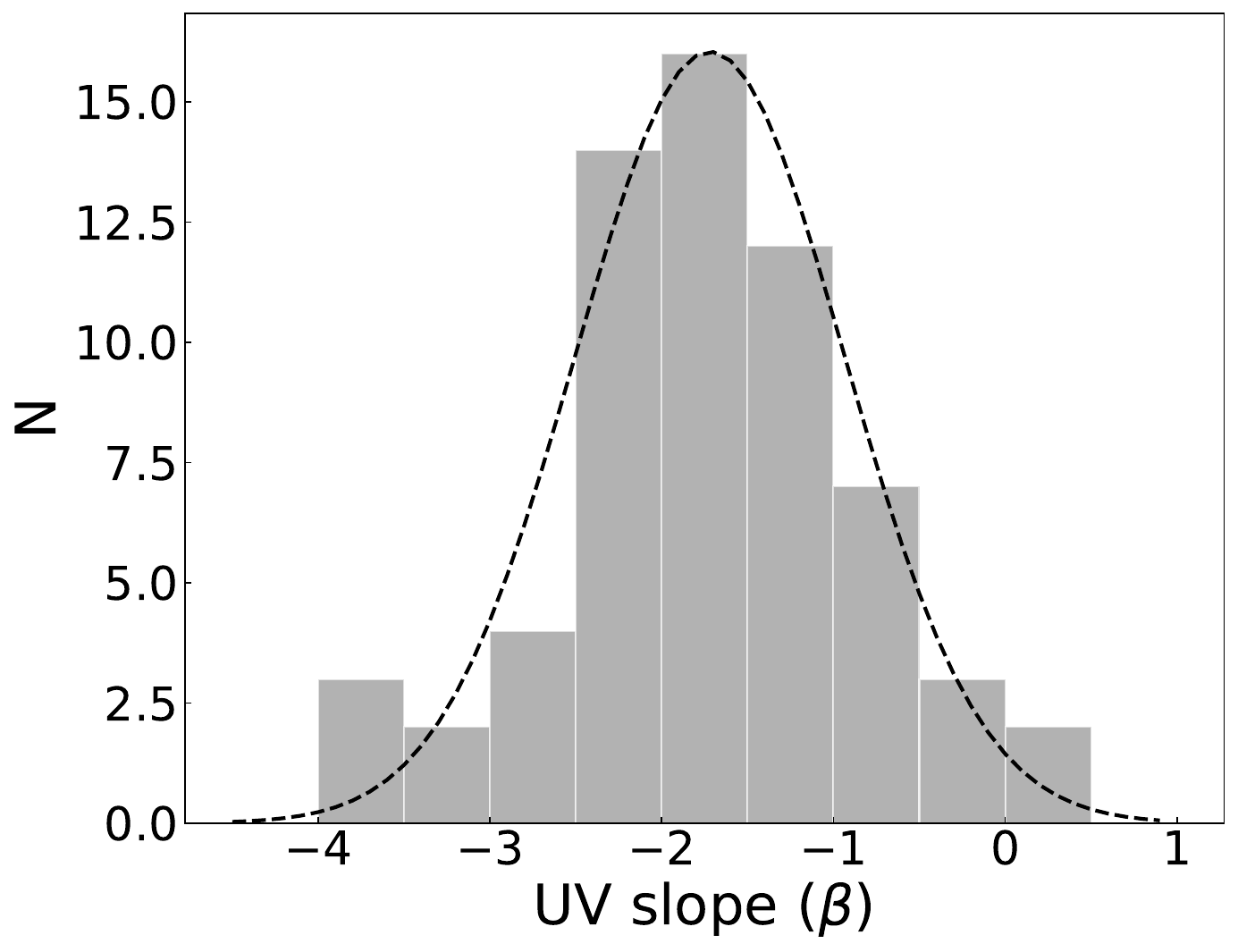}
\caption{Distribution of the UV slope $\beta$ for $z\sim3.7$ LAE sample. The dashed line represents the best-fit Gaussian distribution with an average $\beta=-1.72$ and $\sigma_{\beta}=0.80$.
\label{fig:z37BetaDistribution}}
\end{figure}

Based on the fitting results, the UV magnitude at $\rm 1500 \, \AA$, Ly$\alpha$ line flux, Ly$\alpha$ luminosity,  and the rest-frame Ly$\alpha$ $\rm EW$ ($\rm EW_0$) are measured. Tables \ref{tab:z37table} and \ref{tab:z48table} present the results for the $z\sim3.7$ and $z\sim4.8$ LAEs, respectively. The uncertainties are estimated using the Monte Carlo method. For each LAE, we simulate 10000 mock sources whose broadband and narrowband magnitudes follow the observed magnitudes and error distributions. The $1\sigma$ uncertainties of the UV continuum and Ly$\alpha$ line quantities are determined from the $16^{\rm th}$ and $84^{\rm th}$ percentiles of the cumulative distributions. Photometry of a few LAEs is severely affected by nearby objects, so their properties are not measured and they are excluded in our analyses. Figure \ref{fig:z37BetaDistribution} shows the distribution of the UV slopes $\beta$ for the $z\sim3.7$ LAEs. Our selection criteria are mainly based on the Ly$\alpha$ EW but not sensitive to $\beta$, so the $\beta$ distribution suffers little selection effect and can be regarded as an intrinsic distribution for LAEs at this redshift. By fitting a Gaussian function to the $\beta$ distribution, we obtain an average $\beta=-1.72$ with a scatter of 0.80.

\section{Ly$\alpha$ luminosity function} \label{sec:laeLF}

In this section, we calculate Ly$\alpha$ LFs from the $z\sim3.7$ and $z\sim4.8$ LAEs in our samples. We first estimate sample completeness and correct for selection effects. We then derive the Ly$\alpha$ LFs using the $1/V_{\rm a}$ method and the maximum likelihood method. 

\subsection{Sample completeness}

\begin{figure}[t]
\epsscale{0.9}
\plotone{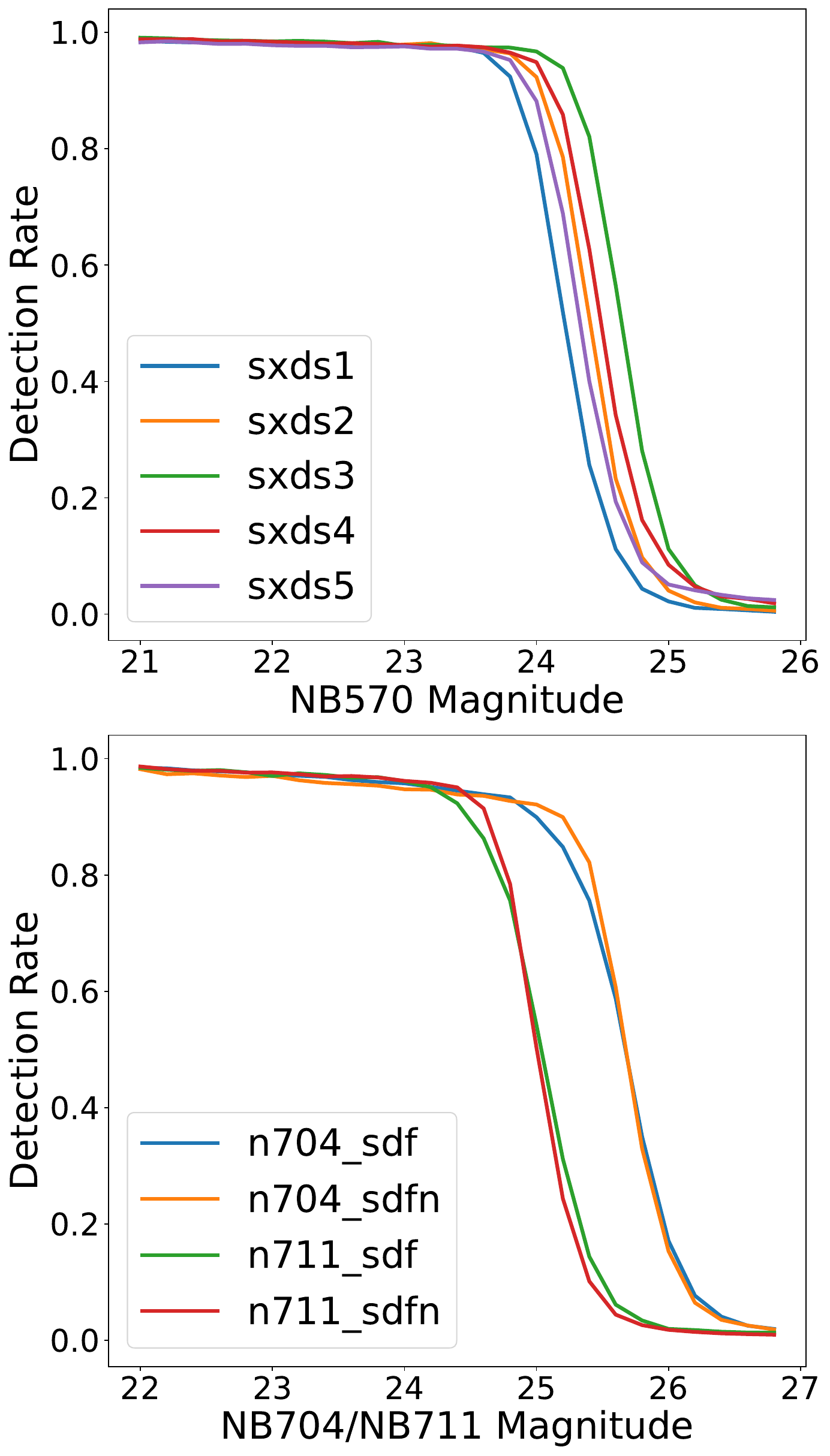}
\caption{Source detection rates in the 5 SXDS subfields (upper panel) and SDF and SDFn (lower panel) as a function of the narrowband magnitudes. Differences between the curves reflect different depths of the narrowband images.
\label{fig:DetectionRate}}
\end{figure}

Incompleteness was introduced into our samples from several sources. The first one is the object detection in the images. Brighter LAEs in the narrow band have higher probabilities to be detected, while some faint LAEs can be missed in this step. We conduct Monte Carlo simulations to estimate object detection probabilities as a function of the narrowband magnitude. We first co-add narrowband images of the spectroscopically confirmed LAEs to obtain typical LAE morphology and sizes in different subfields. We then generate 2500 mock LAEs with different magnitudes that cover our samples, and randomly distribute them into the narrowband images. Finally, we detect these mock LAEs using SExtractor with the same configuration parameters as we did previously, and estimate the detection rate as a function of magnitude. Figures \ref{fig:DetectionRate} shows the results for all subfields. These results will be used to correct detection incompleteness later. 

\begin{figure*}[t]
\epsscale{1}
\plotone{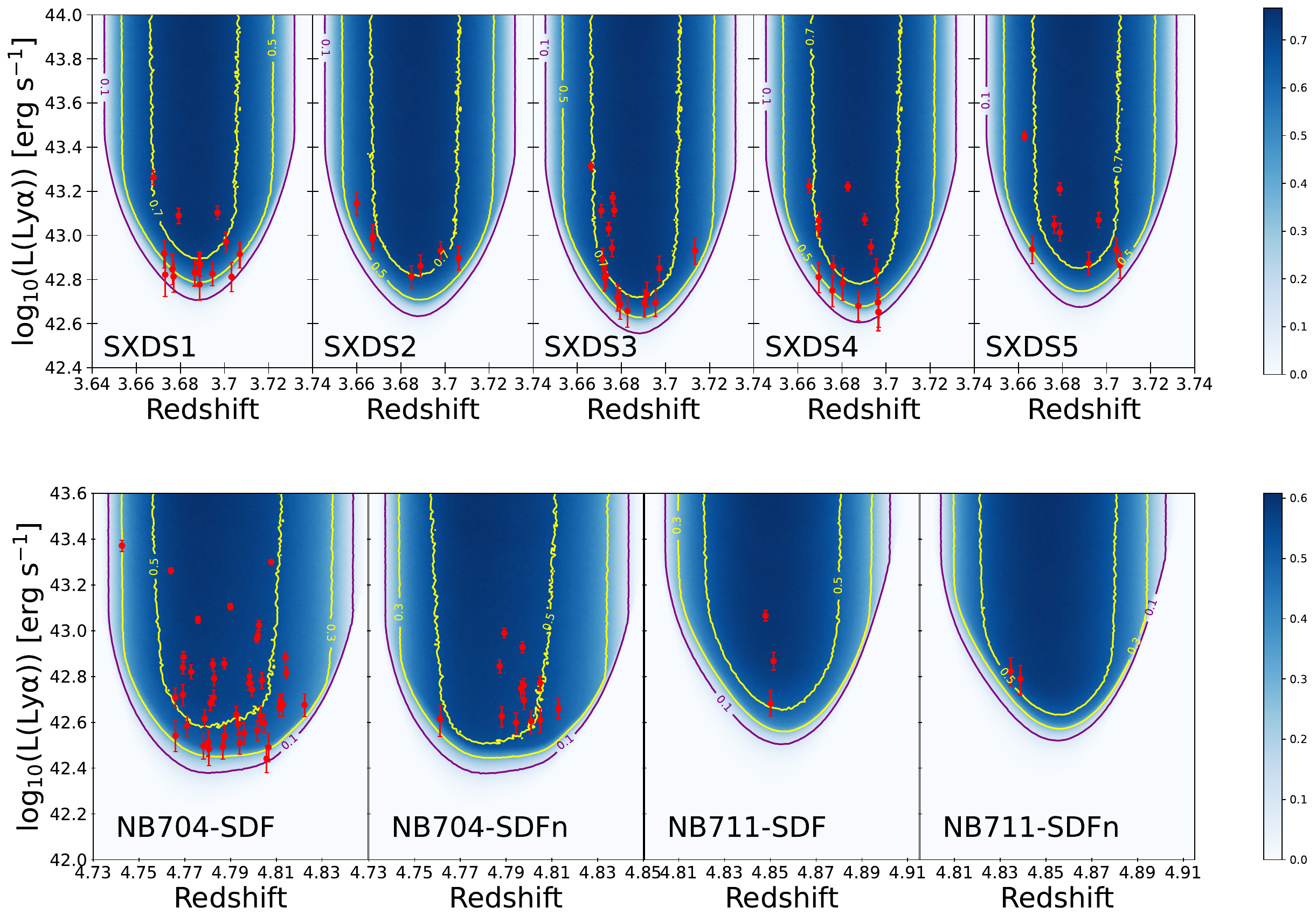}
\caption{The total completeness fraction (combining source detection, target selection, and spectroscopic observations) in different subfields as a function of Ly$\alpha$ luminosity and redshift. The three contours represents the fractions of $10\%$, $50\%$, and $70\%$ in the upper panel, and $10\%$, $30\%$, and $50\%$ in the lower panel. The red points with error bars represent the LAEs in our samples. 
\label{fig:TotCom}}
\end{figure*}

The second source of the sample incompleteness comes from our LAE candidate selection criteria. For example, the color criterion $VR - \rm NB570 > 0.9$ for $z\sim3.7$ LAE candidates roughly corresponds to the rest-frame Ly$\alpha ~ \rm EW_0 \gtrsim 20 \, \AA$, so  LAEs with lower $\rm EW_0$ have lower probabilities be selected. In addition, LAEs with relatively low Ly$\alpha$ luminosities and weak UV continuum can have high $\rm EW_0$, but they can be faint in the narrow band and will not satisfy our magnitude cut $\rm NB>7\sigma$. In narrowband surveys so far, very few LAEs with Ly$\alpha$ luminosities lower than $\sim10^{41.5}$ erg/s are found even with deep images \citep{2011A&A...525A.143C}. We carry out Monte Carlo simulations to estimate the selection completeness. We generate a grid of 10000 mock LAE spectra with different Ly$\alpha$ luminosities and redshifts $(\log(L_{\rm Ly\alpha}), z)$ according to equation (\ref{eq:specModel}). The UV slopes $\beta$ and $\rm EW_0$ are randomly chosen from their intrinsic distributions. For $z\sim3.7$ LAEs, we use the $\beta$ distribution from our sample (Gaussian distribution with $\beta=-1.72\pm0.80$) to represent its intrinsic distribution. Note that $\beta$ is not sensitive to the selection criteria. 
For $z\sim4.8$ LAEs, we are not able to calculate $\beta$ from our data, so we use $\beta=-2.17\pm1.57$ from \cite{2017A&A...608A..10H}. For $\rm EW_0$, its distribution can often be fitted by an exponential distribution, $N=N_0 \times \exp(-w/w_0)$ \citep[e.g.,][]{2007ApJ...667...79G} with a scale length $w_0$. It is usually difficult to obtain the intrinsic $\rm EW_0$ distribution. We use the results from \cite{2017A&A...608A..10H} ($w_0=113\rm \, \AA$ and $w_0=68\rm \, \AA$ for $z\sim3.7$ and $z\sim4.8$ LAEs, respectively), whose LAE sample is not selected by the narrowband technique but by a blind Integral Field Unit (IFU) spectroscopic survey. Their sample reaches a great depth for Ly$\alpha$ luminosity without strong selection effects on $\rm EW_0$, so it can be used to represent the intrinsic $\rm EW_0$ distribution. After applying the IGM absorption to the mock spectra, we convolve them with broad- and narrowband filter transmission curves to obtain their magnitudes as if they were actually observed. Errors for magnitudes are estimated based on the magnitude-error relation of each band in each subfield. Then, we select these mock LAEs with our selection criteria and calculate the selection rate as a function of Ly$\alpha$ luminosity and redshift.

Third, our spectroscopic observations and target identification also brought incompleteness to our samples. As previously mentioned, due to the fiber collision and other constraints, 99 out of 111 LAE candidates at $z\sim3.7$ and 108 out of 151 LAE candidates at $z\sim4.8$ were assigned with fibers. This incompleteness is independently of luminosity. As we also mentioned earlier, our strategy ensures that each real LAE gets enough exposure time for line identification, so we assume that target identification did not introduce any incompleteness. Figure \ref{fig:TotCom} shows the total completeness (combination of object detection, target selection, and spectroscopic observations) in different subfields for the $z\sim3.7$ and $z\sim4.8$ LAE samples. We do not apply any completeness cut when we calculate the Ly$\alpha$ LF in the next two subsections.

\begin{figure}[t]
\epsscale{1.15}
\plotone{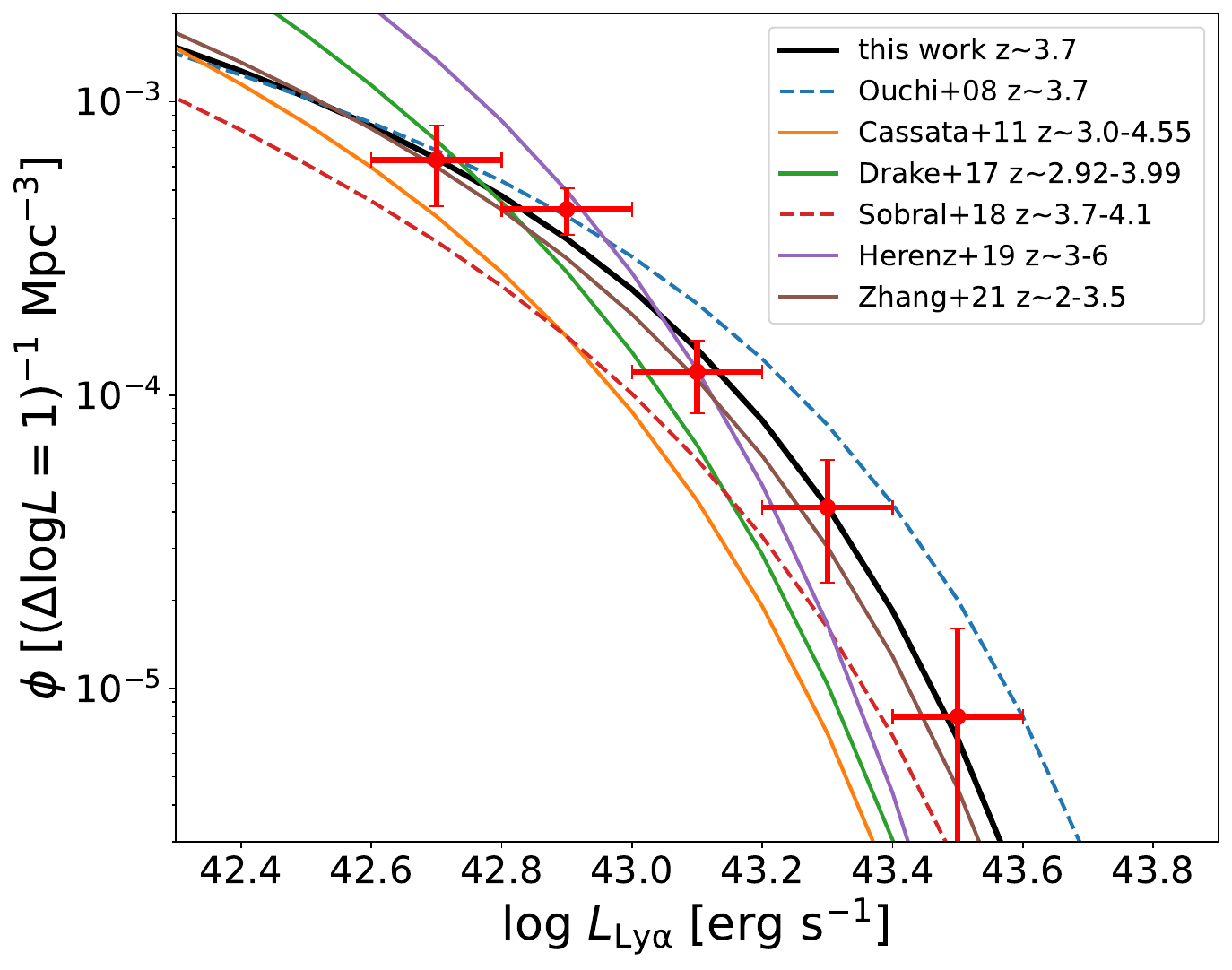}
\caption{Ly$\alpha$ LFs at $z\sim3.7$. The red points with error bars represent the binned LF of the $z\sim3.7$ LAE sample. The vertical error bars denote the $1\sigma$ Poisson error and the horizontal bars indicate the bin size of $\Delta \log L = 0.2$ dex. The black solid line is the best-fit LF with the Schechter function and $\alpha$ fixed to --1.5 for the $z\sim3.7$ LAE sample. Ly$\alpha$ LFs at similar redshifs in the literature are also shown (Ouchi+08: \cite{2008ApJS..176..301O}; Cassata+11: \cite{2011A&A...525A.143C}; Drake+17: \cite{2017A&A...608A...6D}; Sobral+18: \cite{2018MNRAS.476.4725S}; Herenz+19: \cite{2019A&A...621A.107H}; Zhang+21: \cite{2021ApJ...922..167Z}). Dashed and solid lines indicate Ly$\alpha$ LFs calculated from photometric LAE samples and spectroscopically confirmed LAE samples, respectively.
\label{fig:z37LF}}
\end{figure}

\subsection{The $1/V_a$ method}

We calculate the non-parametric binned Ly$\alpha$ LF of our $z\sim3.7$ and $z\sim4.8$ LAEs using the $1/V_a$ method \citep{1980ApJ...235..694A}, a modified version of the $1/V_{\rm max}$ method. $V_a$ is the available comoving volume that a LAE with Ly$\alpha$ luminosity $L$ can be detected in our sample. If we use $p(\log L, z)$ to denote the sample completeness as a function of Ly$\alpha$ luminosity and redshift, then $V_a$ can be expressed as
\begin{equation}
V_a = \omega \int_{\rm z_{min}}^{\rm z_{max}} p(\log L,z) \frac{dV_c}{dz} dz , \label{eq:VaSingle}
\end{equation}
where $\omega$ is the area of the field, $z_{\rm min}$ and $z_{\rm max}$ are determined by the narrowband filters, and $V_c$ is comoving volume. The binned LF is given by
\begin{equation}
\phi(\log L) = \frac{1}{\Delta \log L} \sum_i \frac{1}{V_{a,i}}, \label{eq:binLFSingle}
\end{equation}
where $V_{a,i}$ is $V_a$ for the $i^{th}$ LAE in the sample and $\Delta \log L$ is the bin size. The $1 \sigma$ error for each bin is given by the Poisson statistics, 
\begin{equation}
\sigma(\phi(\log L)) = \frac{1}{\Delta \log L} (\sum_i (\frac{1}{V_{a,i}})^2)^{1/2}. \label{eq:binLFSingleError}
\end{equation} 

\begin{figure}[t]
\epsscale{1.15}
\plotone{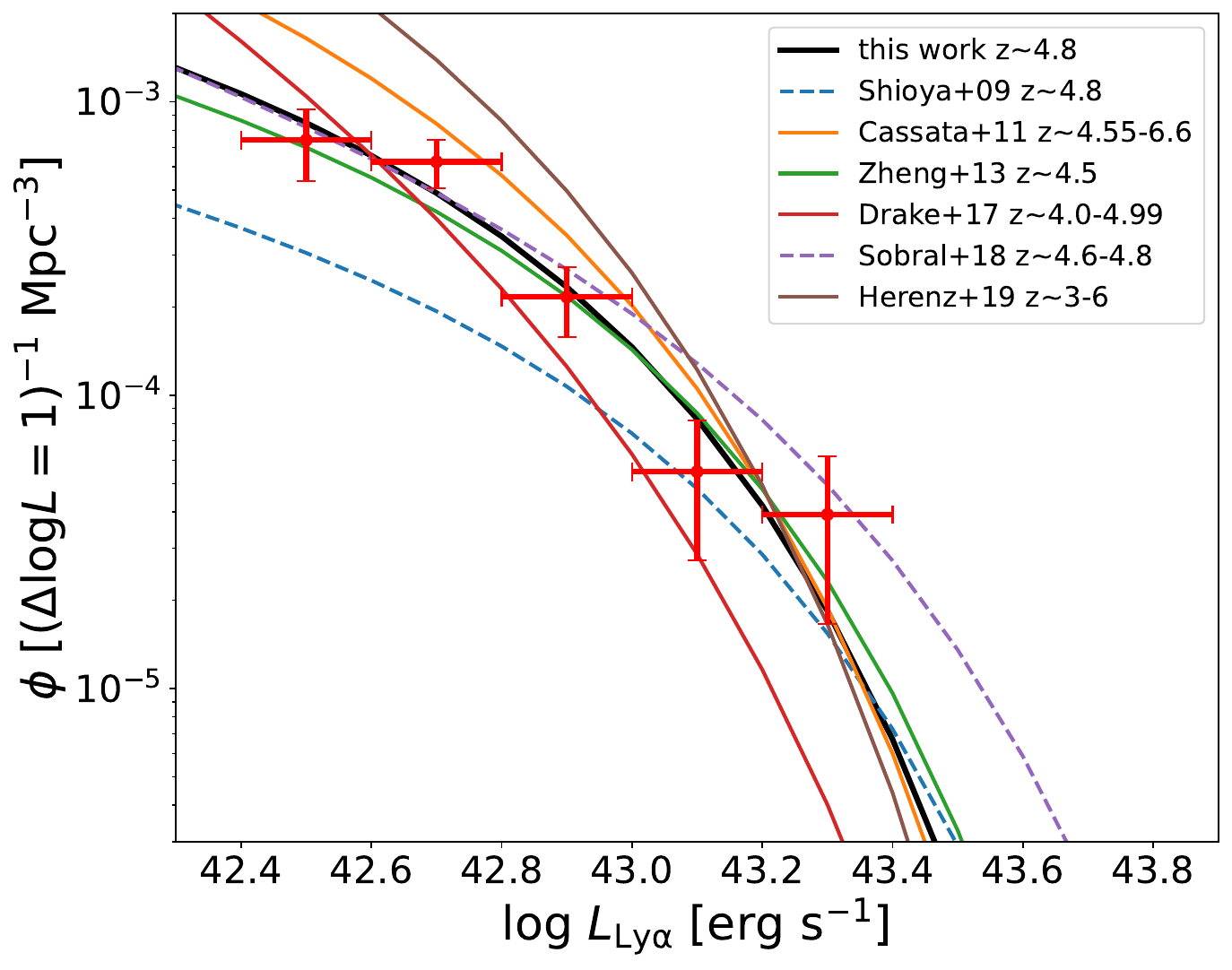}
\caption{Ly$\alpha$ LFs at $z\sim4.8$. The red points with error bars represent the binned LF of the $z\sim4.8$ LAE sample. The vertical error bars denote the $1\sigma$ Poisson error and the horizontal bars indicate the bin size of $\Delta \log L = 0.2$ dex. The black solid line is the best-fit LF with the Schechter function and $\alpha$ fixed to --1.5 for the $z\sim4.8$ LAE sample. Ly$\alpha$ LFs at similar redshifs in the literature are also shown (Shioya+09: \cite{2009ApJ...696..546S}; Cassata+11: \cite{2011A&A...525A.143C}; Zheng+13: \cite{2013MNRAS.431.3589Z}; Drake+17: \cite{2017A&A...608A...6D}; Sobral+18: \cite{2018MNRAS.476.4725S}; Herenz+19: \cite{2019A&A...621A.107H}). Dashed and solid lines indicate Ly$\alpha$ LFs calculated from photometric LAE samples and spectroscopically confirmed LAE samples, respectively.
\label{fig:z48LF}}
\end{figure}

If several subfields are considered, equation \ref{eq:VaSingle} can be written as
\begin{equation}
V_a = \sum_j \omega_j \int_{\rm z_{min}}^{\rm z_{max}} p_j(\log L,z) \frac{dV_c}{dz} dz , \label{eq:VaMulti}
\end{equation}
where $\omega_j$ is the area of the $j^{th}$ subfield, and $p_j(\log L,z)$ is the completeness for the $j^{th}$ subfield. The effective areas for all subfields are given in Section 2.3. The maximum available comoving volume is $\rm \sim 6 \times 10^5 \, Mpc^3$ for our $z\sim3.7$ LAE sample and $\rm \sim 4 \times 10^5 \, Mpc^3$ for our $z\sim4.8$ LAE sample. The red points in Figures \ref{fig:z37LF} and \ref{fig:z48LF} show the binned Ly$\alpha$ LF. The vertical error bars represent Poisson errors and the horizontal bars indicate bin sizes. 

\subsection{Maximum likelihood method}

To parameterize the Ly$\alpha$ LF, we use the Schechter function \citep{1976ApJ...203..297S},
\begin{equation}
\phi(\log L) \mathrm{d}\log L = \ln 10 \times \phi^* \left (\frac{L}{L^*}\right)^{\alpha+1} \exp \left (-\frac{L}{L^*} \right) \mathrm{d} \log L , \label{eq:ScheLF}
\end{equation}
where $\phi^*$ is the characteristic volume density, $L^*$ is the characteristic luminosity, and $\alpha$ is the faint-end slope. We do not fit the Schechter function to the binned Ly$\alpha$ LF because the chosen size of the luminosity bin may potentially affect the fitting results. Instead, we adopt a maximum likelihood method \citep[e.g.,][]{2013ApJ...769...83C} to get the best-fit Schechter function parameters. 

Given ($\phi^*_{\rm Ly\alpha}, L^*_{\rm Ly\alpha}, \alpha$), the expected number of LAEs with $L < L({\rm Ly\alpha}) < L+\Delta L$ that should be discovered in our sample is denoted by $\lambda = f(L) \Delta L$, where
\begin{equation}
f(L)=\sum_j \omega_j \int_{\rm z_{min}}^{\rm z_{max}} p_j(\log L,z) \phi(L) \frac{dV_c}{dz} dz . \label{eq:probL}
\end{equation}
The actual number $n$ of LAEs found with $L < L({\rm Ly\alpha}) < L+\Delta L$ follows Poisson statistics with $P(n)=\frac{e^{-\lambda} \lambda^n}{n!}$. If we reduce $\Delta L$ enough so that in each luminosity bin either 0 LAE or 1 LAE is found, since $P(1)= \lambda e^{-\lambda}$ and $P(0)=e^{-\lambda}$, the likelihood function based on our sample can be written as 
\begin{equation}
P=\prod_i \left (e^{-f(L_i)\Delta L} f(L_i) \Delta L \right) \times \prod_{k \neq i} e^{-f(L_k)} , \label{eq:likelihood}
\end{equation}
where $L_i$ is the Ly$\alpha$ luminosity for each confirmed LAE and $L_k$ refers to luminosity bin without any discovered LAE. After taking small enough $\Delta L$, the logarithm of the likelihood function can be expressed by
\begin{equation}
    \begin{aligned}
        \ln P &= -\sum_j \left ( \omega_j \int_{z_{\rm min}}^{\rm z_{max}} \int_0^{+\infty}p_j(L,z)\phi(L)dL \frac{dV_c}{dz}dz \right) \\ 
        &+ \sum_i \ln \left( \sum_j \omega_j \int_{\rm z_{min}}^{\rm z_{max}} p_j(L_i, z) \phi(L_i) \frac{dV_c}{dz} dz \right),  \label{eq:loglikelihood}
    \end{aligned}
\end{equation}
where $i$ represents the $i^{\rm th}$ LAE and $j$ represents the $j^{\rm th}$ subfield. In this formula, all three Schechter function parameters $\phi^*_{\rm Ly\alpha}, L^*_{\rm Ly\alpha}$, and $\alpha$ are free parameters. By varying ($\phi^*_{\rm Ly\alpha}, L^*_{\rm Ly\alpha}, \alpha$), we find the best-fit parameters that maximize $\ln P$, or minimize $S=-2 \times \ln P$. 

\begin{deluxetable}{cccc} 
\renewcommand\arraystretch{1.3}
\tabletypesize{\footnotesize}
\tablewidth{0pt}
\tablenum{4}
\tablecaption{Best-fit parameters of the Schechter function for the $z\sim3.7$ and 4.8 Ly$\alpha$ LFs. \label{tab:SchPara}}
\tablehead{
\colhead{Redshift} & \colhead{$\alpha$} & \colhead{$\rm \log_{10}\,$$L_{\rm Ly\alpha}^*$} & 
\colhead{$\rm \log_{10}\,$$\phi_{\rm Ly\alpha}^*$}\\
\colhead{} & \colhead{} & \colhead{($\rm erg$ $\rm s^{-1}$)} & 
\colhead{($\rm Mpc^{-3}$)}
}
\startdata
\hline
$z\sim3.7$ & $-$1.5 (fixed) & $\rm 42.86^{+0.10}_{-0.07}$ & $\rm -3.34^{+0.14}_{-0.19}$\\
  & $-$1.6 (fixed) & $\rm 42.89^{+0.10}_{-0.08}$ & $\rm -3.38^{+0.14}_{-0.20}$\\
  & $-$1.7 (fixed) & $\rm 42.92^{+0.11}_{-0.08}$ & $\rm -3.43^{+0.15}_{-0.21}$\\
  & $-$1.8 (fixed) & $\rm 42.95^{+0.11}_{-0.09}$ & $\rm -3.49^{+0.16}_{-0.23}$\\
  & $-$1.9 (fixed) & $\rm 42.98^{+0.13}_{-0.09}$ & $\rm -3.55^{+0.16}_{-0.25}$\\
  & $-$2.0 (fixed) & $\rm 43.02^{+0.13}_{-0.10}$ & $\rm -3.63^{+0.18}_{-0.26}$\\
\hline
$z\sim4.8$ & $-$1.5 (fixed) & $\rm 42.76^{+0.10}_{-0.08}$ & $\rm -3.33^{+0.13}_{-0.18}$\\
 & $-$1.6 (fixed) & $\rm 42.79^{+0.12}_{-0.08}$ & $\rm -3.38^{+0.14}_{-0.20}$\\
 & $-$1.7 (fixed) & $\rm 42.83^{+0.08}_{-0.12}$ & $\rm -3.44^{+0.20}_{-0.14}$\\
 & $-$1.8 (fixed) & $\rm 42.86^{+0.14}_{-0.09}$ & $\rm -3.51^{+0.15}_{-0.24}$\\
 & $-$1.9 (fixed) & $\rm 42.90^{+0.15}_{-0.10}$ & $\rm -3.59^{+0.16}_{-0.27}$\\
 & $-$2.0 (fixed) & $\rm 42.95^{+0.17}_{-0.11}$ & $\rm -3.69^{+0.18}_{-0.29}$\\
\enddata
\end{deluxetable}

Because our samples are not deep enough to constrain the faint-end slope $\alpha$ of the Ly$\alpha$ LF, we fix $\alpha$ from $-1.5$ to $-2.0$ in a step of 0.1 and then fit the other two parameters. Table \ref{tab:SchPara} shows the best-fit results. The $1 \sigma$ error for each parameter is determined by marginalizing over the other parameter and finding the $16^{\rm th}$ and $84^{\rm th}$ percentiles of the cumulative distribution. In Figures \ref{fig:z37LF} and \ref{fig:z48LF}, we show the Ly$\alpha$ LFs with $\alpha=-1.5$ that was commonly used in previous studies \citep[e.g.,][]{2007ApJ...667...79G, 2008ApJS..176..301O, 2022ApJ...926..230N}. The best-fit Schechter functions are consistent with the binned LF, which suggests that the Ly$\alpha$ LF of LAEs is approximated well by the Schechter function.

In the above procedure, we did not apply any luminosity cut when we derived the LFs. In addition, the Eddington bias is small compared to the uncertainties of the LFs. We estimate the effect of the Eddington bias using Monte Carlo simulations below. For each LAE, we randomly choose a Ly$\alpha$ luminosity that follows the observed Gaussian distribution of its Ly$\alpha$ luminosity. We then derive the LFs at the two redshifts with $\alpha$ fixed to $-1.5$, as we did earlier. We repeat the simulation 100 times and find that the LF results are all consistent with the previous results within one sigma uncertainties. Therefore, we did not consider the Eddington bias above.

\section{Discussion} \label{sec:discussion}

\subsection{Comparison with previous studies}

In Figures \ref{fig:z37LF} and \ref{fig:z48LF}, Ly$\alpha$ LFs at similar redshifts from the literature are overploted. These studies use different methods (including the narrowband technique, IFU spectroscopy, and slit spectroscopy) to obtain LAE samples, so their sample sizes, survey volumes, redshift ranges, and Ly$\alpha$ luminosity limits are different. Nevertheless, our Ly$\alpha$ LFs at the two redshifts are roughly consistent with most of the previous measurements. 

We first compare Ly$\alpha$ LFs at $z\sim3.7$. \cite{2008ApJS..176..301O} used the same narrowband filter NB570 and selected LAEs in the same field SXDS as we did. They measured their Ly$\alpha$ LF based on the photometric sample (they fixed $\alpha=-1.5$). Their LF at the faint end agrees well with our faint-end Ly$\alpha$ LF, but their bright-end LF is a factor of $2-3$ higher than ours. The reason is unclear, since bright LAEs are relatively easy to detect and identify, with little selection bias or incompleteness. On the other hand, our spectroscopically confirmed sample ruled out contaminants that may exist in previous photometric samples. The Ly$\alpha$ LF from \cite{2018MNRAS.476.4725S} is systematically lower that ours from the faint end to the bright end, possibly because they used a medium band ($\rm FWHM \sim 280 \, \AA$) to select LAE candidates with large Ly$\alpha$ EWs ($\rm EW_0 > 50 \, \AA$). The \cite{2011A&A...525A.143C} LAE sample was obtained using slit spectroscopy and the \cite{2017A&A...608A...6D}  sample was obtained using MUSE IFU spectroscopy. Their Ly$\alpha$ LFs are not consistent with each other, and both are much lower than ours at the bright end. It is likely because their survey volumes are relatively small ($\rm \lesssim 10^5 \, Mpc^3$) and thus there are not enough bright LAEs. At the faint end, Ly$\alpha$ LFs from \cite{2017A&A...608A...6D} and \cite{2019A&A...621A.107H} are higher than our result. As \cite{2019A&A...621A.107H} pointed out, LAEs exhibit diffuse extended low surface-brightness halos. If taking this into consideration in the calculation of the Ly$\alpha$ luminosity, \cite{2019A&A...621A.107H} found that the Ly$\alpha$ LFs at the faint end are underestimated in the narrowband studies. We, like previous narrowband studies, did not take this into account. The \cite{2021ApJ...922..167Z} LAE sample is from a large HETDEX IFU spectroscopic survey of bright LAEs, and their Ly$\alpha$ LF at $z\sim2-3.5$ is consistent with (or slightly below) our LF. The consistency may reflect the fact that there is a moderate increase of Ly$\alpha$ LF from $z\sim2$ to 3 and almost no evolution from $z\sim3$ to 6 \citep[e.g.,][]{2016ApJ...823...20K}.

We then compare $z\sim4.8$ Ly$\alpha$ LFs. We first compare our LF with \cite{2009ApJ...696..546S} who used the same narrowband filter NB711 as we did to obtain a photometric LAE sample in the COSMOS field. Their Ly$\alpha$ LF is similar to ours at the bright end, but significantly smaller at the faint end. The reason is likely that \cite{2009ApJ...696..546S} derived their Ly$\alpha$ LF without considering completeness correction, because a sample suffers larger incompletenesses at fainter luminosities. 
Our $z\sim4.8$ LF agrees well with the $z\sim4.5$ Ly$\alpha$ LF in \cite{2013MNRAS.431.3589Z}, which used a spectroscopically confirmed LAE sample and incorporated the LAE sample from \cite{2007ApJ...671.1227D}. The Ly$\alpha$ LF by \cite{2017A&A...608A...6D} based on the MUSE data is smaller at the bright end due to their smaller survey volume. The Ly$\alpha$ LFs by \cite{2011A&A...525A.143C} based on the slit spectroscopy and by \cite{2019A&A...621A.107H} based on the MUSE data are similar to our LF at the bright end (because at $z\sim4.8$ our survey volume is comparable to theirs) and higher than our Ly$\alpha$ LF at the faint end. This is likely because their slit spectroscopy and IFU spectroscopy data probed fainter Ly$\alpha$ luminosities. In addition,  \cite{2019A&A...621A.107H} took diffuse low surface brightness halos into consideration, as we explained earlier. 

\begin{figure}[t]
\plotone{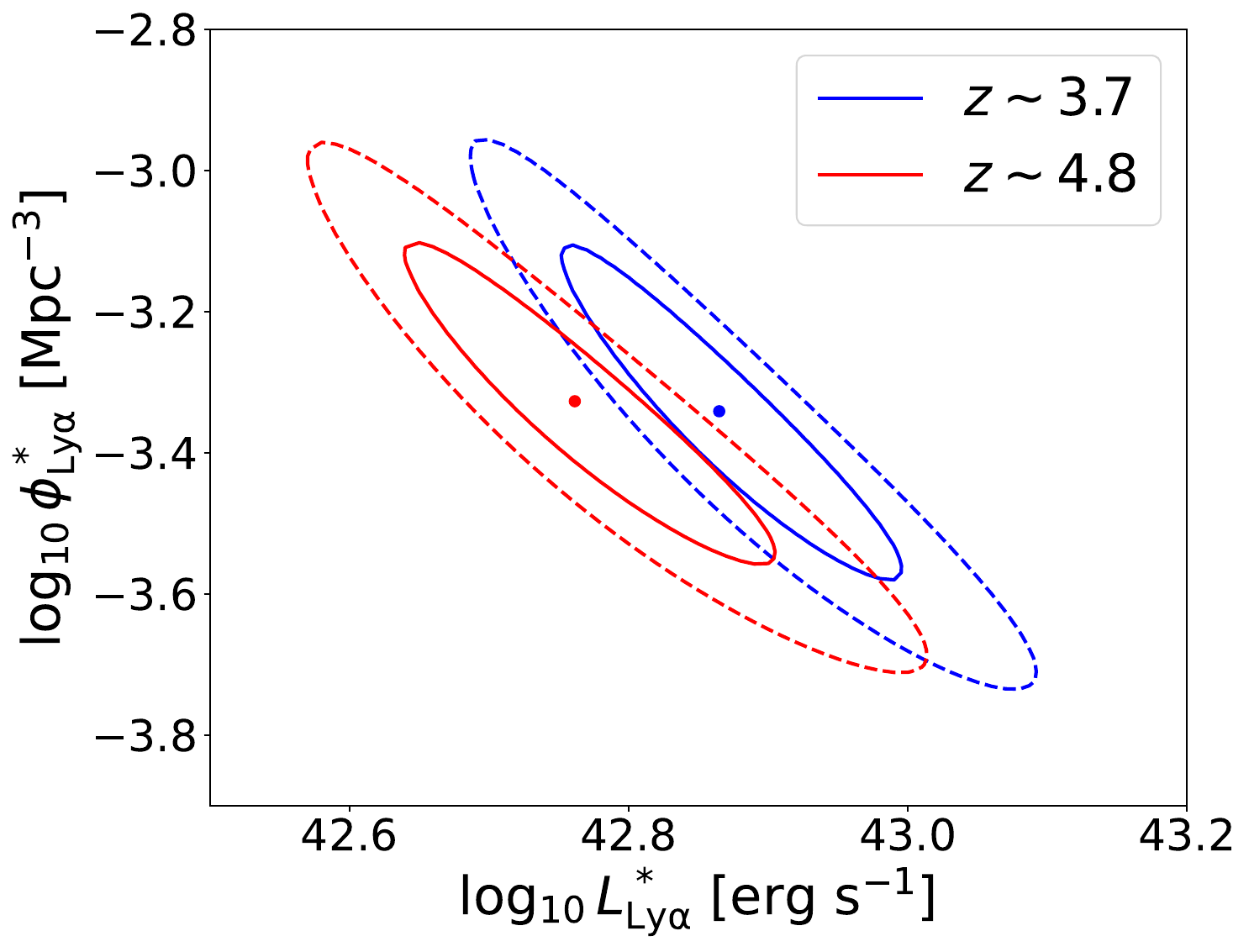}
\caption{Likelihood contours in the $\phi^*_{\rm Ly\alpha} - L^*_{\rm Ly\alpha}$ parameter space for the two LAE samples at  $z\sim3.7$ (blue) and $z\sim4.8$ (red). The blue and red dots show the best-fit $\phi^*_{\rm Ly\alpha}$ and $L^*_{\rm Ly\alpha}$. The solid and dashed lines represent $1\sigma \, \rm (68\%)$ and $2\sigma \, \rm (95\%)$ confidence regions, respectively.
\label{fig:LFContour}}
\end{figure}

\subsection{Evolution of the Ly$\alpha$ LF}

We further study the evolution of the LF parameters. In Figure \ref{fig:LFContour}, we plot the likelihood contours in the $\phi^*_{\rm Ly\alpha} - L^*_{\rm Ly\alpha}$ space derived from the $z\sim3.7$ and 4.8 LFs with fixed $\alpha =-1.5$. The $1\sigma \, \rm (68\%)$ and $2\sigma \, \rm (95\%)$ confidence regions are determined by respectively setting $\Delta S=2.30$ and $\Delta S=6.17$ when fitting two free parameters ($\phi^*_{\rm Ly\alpha}$ and $L^*_{\rm Ly\alpha}$) in the maximum likelihood method. Although the $1\sigma$ contours barely overlap, the $2\sigma$ contours overlap with each other. The best-fit $\phi^*$ at the two redshifts are almost the same and the best-fit $L^*$ shows a small decrease ($\sim 0.1$ dex) from $z\sim3.7$ to $z\sim4.8$. Therefore, the Ly$\alpha$ LF evolves little from $z\sim3.7$ to 4.8.

\begin{figure}[t]
\plotone{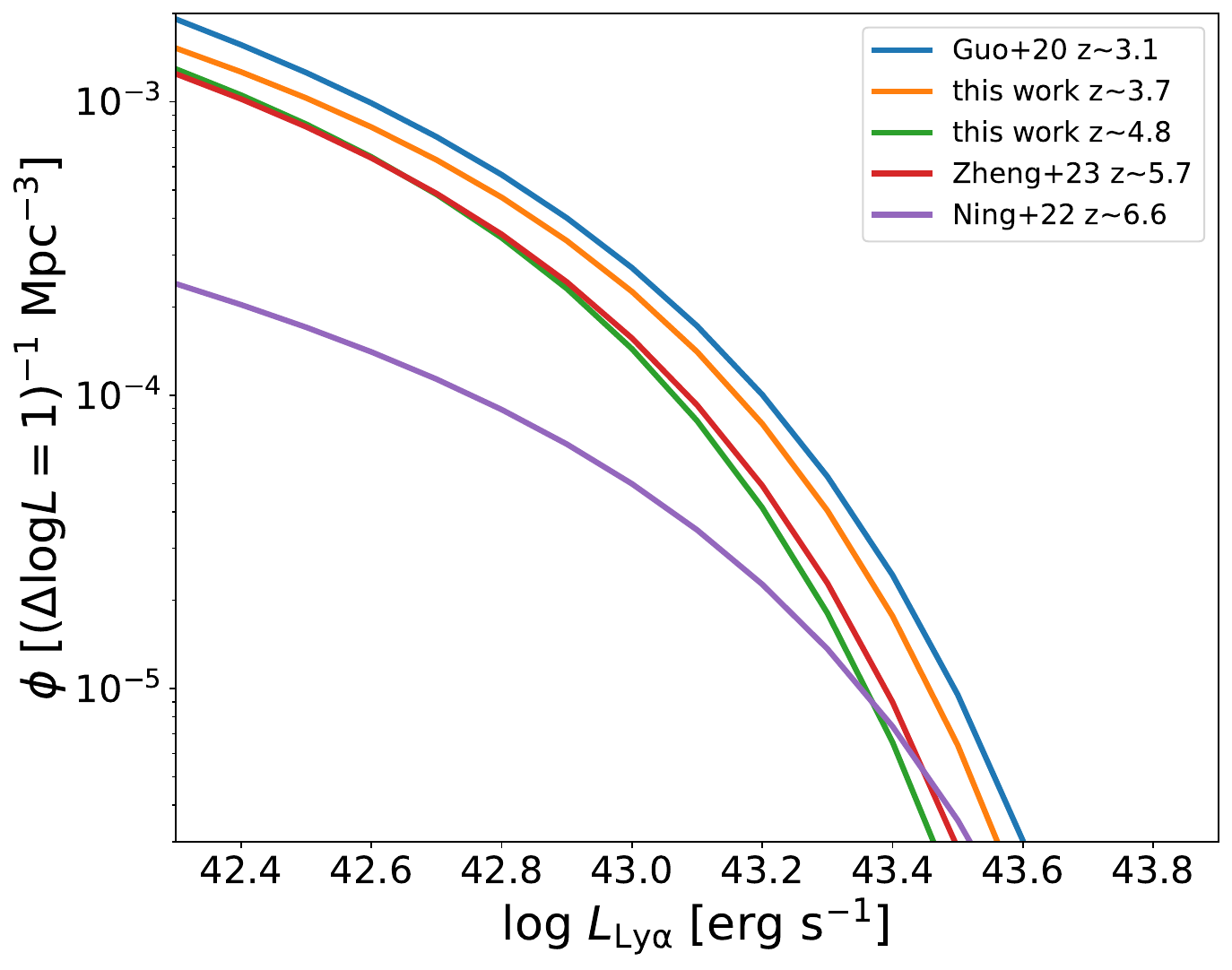}
\caption{Ly$\alpha$ LF at several redshifts based on spectroscopically confirmed LAE samples. $z\sim3.1$: \cite{2020ApJ...902..137G}; $z\sim3.7$: this work; $z\sim4.8$: this work; $z\sim5.7$: Zheng et al. (in preparation); $z\sim6.6$: \cite{2022ApJ...926..230N}. These lines are the best-fit Schechter functions. 
\label{fig:LFComp}}
\end{figure}

In Figure \ref{fig:LFComp}, we plot the Ly$\alpha$ LFs at several redshifts from $z\sim 3.1$ to $z\sim 6.6$, including results from our work, \cite{2020ApJ...902..137G}, \cite{2022ApJ...926..230N}, and Zheng et al. (in preparation, hereafter Z23). \cite{2020ApJ...902..137G} calculated a Ly$\alpha$ LF at $z\sim3.1$ using a large spectroscopic sample of 166 LAEs over $\sim 1.2 \, \rm deg^2$. \cite{2022ApJ...926..230N} derived a Ly$\alpha$ LF at $z\sim6.6$ based on 36 spectroscopically confirmed LAEs over $\sim 2 \, \rm deg^2$. Z23 calculated a Ly$\alpha$ LF at $z\sim5.7$ based on a large sample of 260 spectroscopically confirmed LAEs constructed by \cite{2020ApJ...903....4N}. The imaging data used in these studies were reduced in the same manner as in our work \citep{2013ApJ...772...99J}, the LAEs were selected using the narrowband technique in a consistent way, the areas covered are all large enough to avoid the effect of cosmic variance, and the Ly$\alpha$ LFs were all based on spectroscopically confirmed LAE samples. Therefore, the evolution of the Ly$\alpha$ LF can be measured reliably from these studies. All these studies are based on narrowband selected LAE samples, and are not deep enough to reach faint Ly$\alpha$ luminosities, so $\alpha$ was fixed to about --1.5 in these studies. Figure \ref{fig:LFComp} shows that the Ly$\alpha$ LF evolves slowly from $z\sim3.1$ to 5.7, with a factor of $\sim3$ decrease in $\phi^*_{\rm Ly\alpha}$ and a factor of $\sim1.5$ decrease in $L^*_{\rm Ly\alpha}$. The similar conclusion that the Ly$\alpha$ LF has no significant evolution from $z\sim3$ to $z\sim6$ is also supported by \cite{2008ApJS..176..301O}, \cite{2011A&A...525A.143C}, \cite{2017A&A...608A...6D}, and \cite{2019A&A...621A.107H}. 

In contrary to the slow evolution from $z\sim3.1$ to 5.7, the Ly$\alpha$ LF declines rapidly from $z\sim5.7$ to $z\sim6.6$ at the faint end. Such a decrease can be explained (at least partly) by cosmic reionization. Cosmic reionization is the phase transition of the IGM when the neutral hydrogen (HI) was gradually ionized by ionizing photons at $z\ge6$ \citep{2006ARA&A..44..415F}. The universe became largely ionized after reionization. Therefore, Ly$\alpha$ photons emitted by LAEs at $z\sim6.6$ are more absorbed/scattered (compared with $z\sim5.7$ LAEs) in the IGM, and thus LAEs at $z\sim6.6$ have fainter Ly$\alpha$ luminosities, especially at the faint end. At the bright end, however, the Ly$\alpha$ LF at $z\sim6.6$ seems unaffected by the IGM absorption, probably due to the existence of large ionized bubbles around luminous LAEs that allow Ly$\alpha$ photons to escape \citep[e.g.,][]{2022ApJ...926..230N}.

\section{Summary} \label{sec:summary}

We have presented one of the largest spectroscopically confirmed sample of LAEs at $z\sim3.7$ and $z\sim4.8$. Using the narrowband technique, we selected LAE candidates with deep broadband and narrowband images taken by the Subaru Suprime-Cam: 112 $z\sim3.7$ LAE candidates in SXDS were selected in the narrowband NB570, and 151 $z\sim$ 4.8 LAE candidates in SDF and SDFn were selected in the narrowbands NB704 and NB711. Based on the spectra taken by the MMT Hectospec spectrograph, we finally confirmed 71 $z\sim3.7$ LAEs and 69 $z\sim4.8$ LAEs. 

We determined the Ly$\alpha$ redshifts of the LAEs by fitting a composite Ly$\alpha$ line profile to individual LAE spectra. From secure redshifts and deep broadband and narrowband photometry, we calculated Ly$\alpha$ luminosity, Ly$\alpha$ $\rm EW_0$, and $\rm M_{UV}$ for each LAE. The $z\sim3.7$ LAEs and the $z\sim4.8$ LAEs span Ly$\alpha$ luminosity ranges of $\sim 10^{42.6} - 10^{43.7} \,\rm erg\, s^{-1}$ and $\sim 10^{42.4} - 10^{43.4} \,\rm erg\, s^{-1}$, respectively. They represent the most Ly$\alpha$-luminous galaxies at the two redshifts.

We have derived Ly$\alpha$ LFs at $z\sim3.7$ and 4.8 based on the two LAE samples, after considering sample incompletenesses. The binned Ly$\alpha$ LFs are well approximated by the Schechter function. We determined the best-fit $\phi^*_{\rm Ly\alpha}$ and $L^*_{\rm Ly\alpha}$ by varying the slope $\alpha$ from --1.5 to --2.0 in the maximum likelihood method. The Ly$\alpha$ LFs  show little evolution between the two redshifts. Our Ly$\alpha$ LFs are broadly consistent with previous measurements within a factor of $2-3$. By comparing with Ly$\alpha$ LFs at other redshifts in the literature, we found that Ly$\alpha$ LFs have no significant evolution from $z\sim3.1$ to $z\sim5.7$, but decline rapidly from $z\sim5.7$ to $z\sim6.6$ at the faint end. 

\begin{acknowledgments}

We thank the referee for comments that strengthened this manuscript.
We acknowledge support from
the National Key R\&D Program of China (2022YFF0503401),
the National Science Foundation of China (11721303, 11890693, 12225301),
and the China Manned Space Project with No. CMS-CSST-2021-A05 and  CMS-CSST-2021-A07.
We thank Y. Guo and Y. Ning for helpful discussions.
Observations reported here were obtained at the MMT Observatory, a joint facility of the University of Arizona and the Smithsonian Institution.

\end{acknowledgments}

\vspace{5mm}
\facilities{Subaru (Suprime-Cam), MMT (Hectospec)}

\software{Astropy \citep{2013A&A...558A..33A, 2018AJ....156..123A, 2022ApJ...935..167A}, 
          SExtractor \citep{1996A&AS..117..393B},
          HSRED
          }

\bibliography{ms}{}
\bibliographystyle{aasjournal}

\newpage

\appendix

\begin{figure}[h]
\epsscale{1.15}
\plotone{fig3-1-1.pdf}
\caption{Same as Figure \ref{fig:z37-600}, but for all LAEs at $z\sim3.7$ with $\rm 600 \, lines \, mm^{-1}$ grating spectra.
\label{fig:z37-600-all-1}}
\end{figure}

\begin{figure}[h]
\figurenum{14}
\epsscale{1.15}
\plotone{fig3-1-2.pdf}
\caption{Continued.
\label{fig:z37-600-all-2}}
\end{figure}

\begin{figure}[h]
\epsscale{1.15}
\plotone{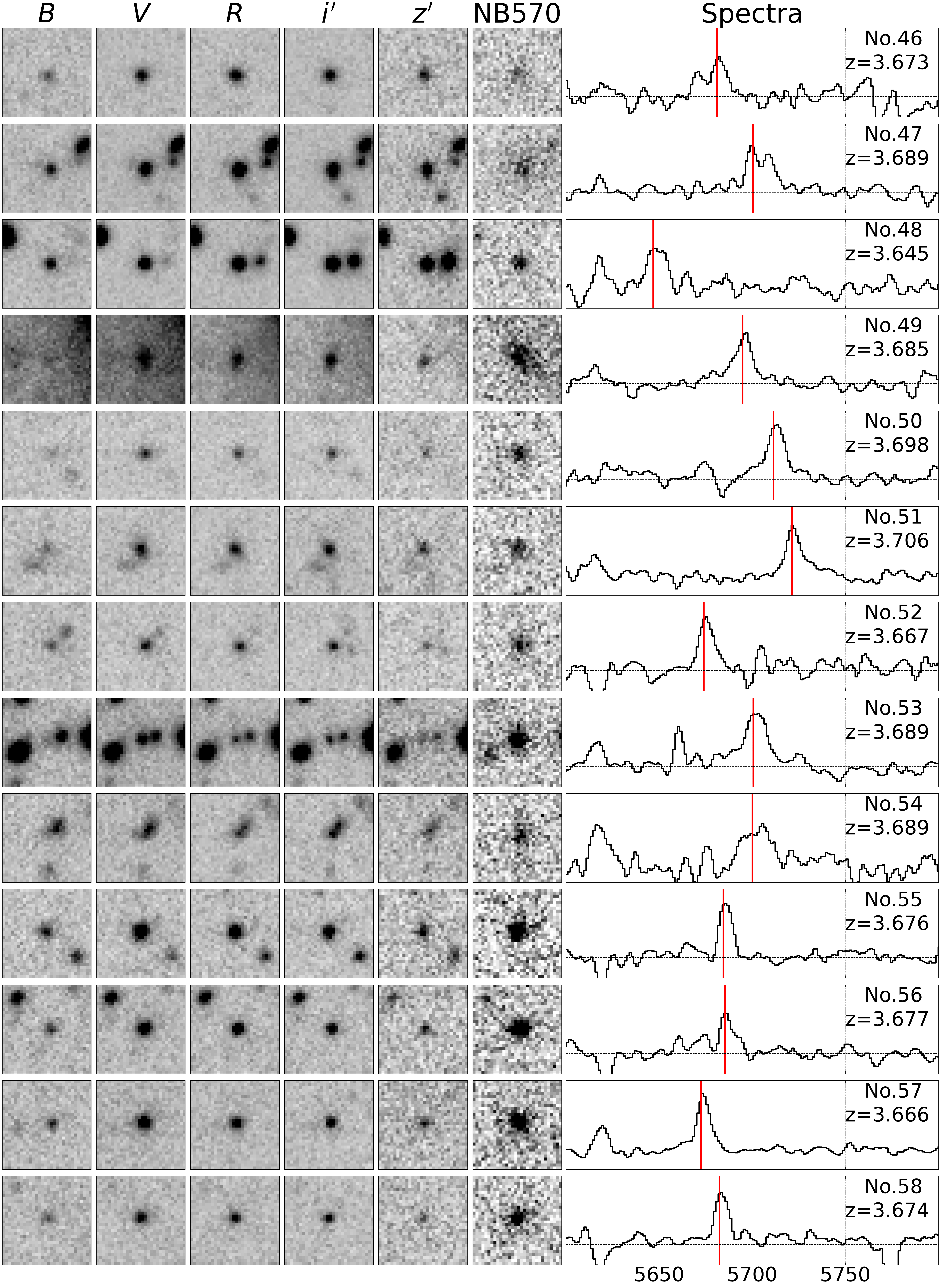}
\caption{Same as Figure \ref{fig:z37-600}, but for all LAEs at $z\sim3.7$ with $\rm 270 \, lines \, mm^{-1}$ grating spectra.
\label{fig:z37-270-all-1}}
\end{figure}

\begin{figure}[h]
\figurenum{15}
\epsscale{1.15}
\plotone{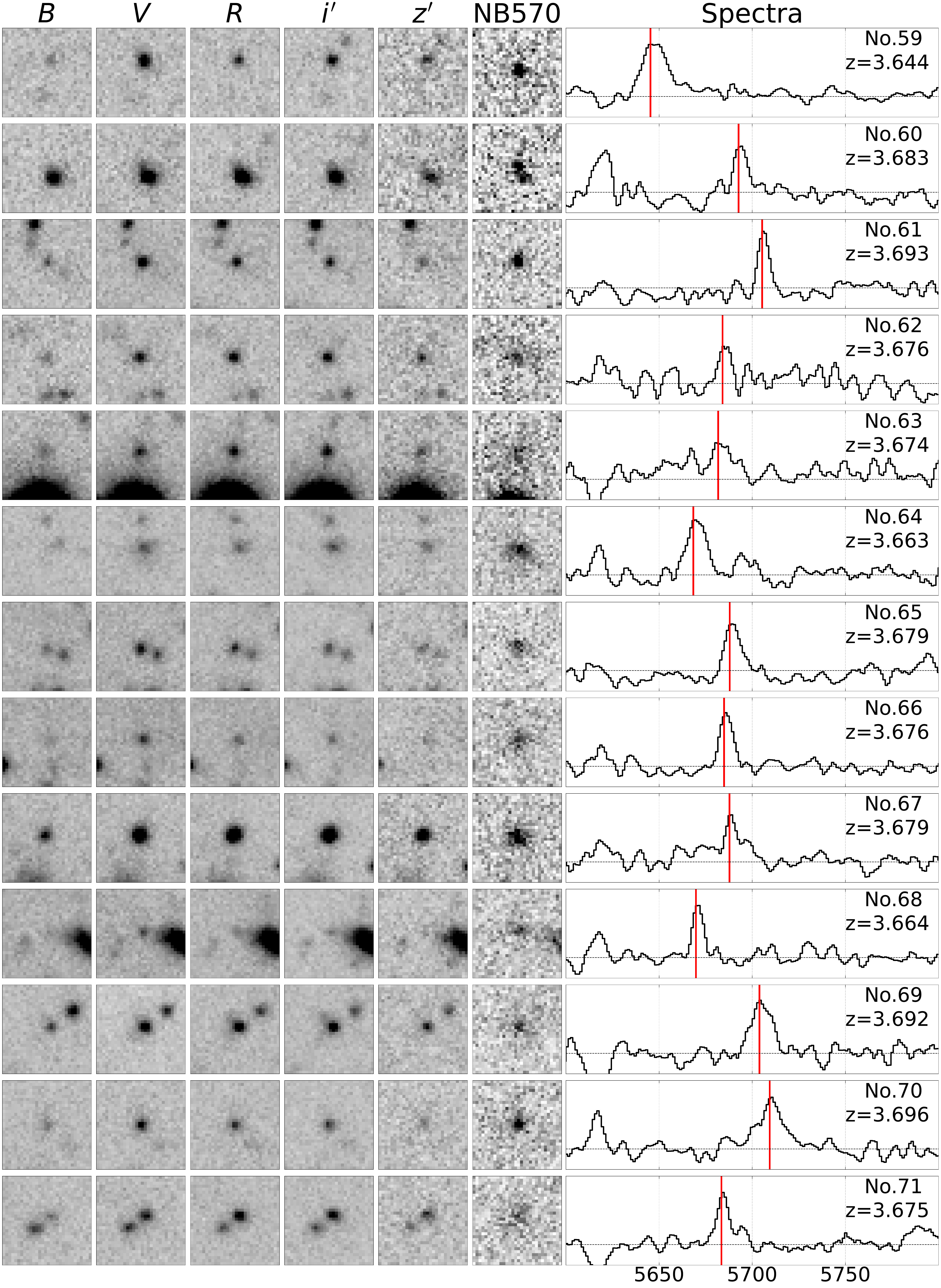}
\caption{Continued.
\label{fig:z37-270-all-2}}
\end{figure}

\newpage

\begin{figure}[h]
\epsscale{1.15}
\plotone{fig4-1-1.pdf}
\caption{Same as Figure \ref{fig:z48-704}, but for all $z\sim4.8$ LAEs selected with narrow band NB704. Objects without $B$, $V$, and $z'$ images are in the SDFn.
\label{fig:z48-704-all-1}}
\end{figure}

\begin{figure}[h]
\figurenum{16}
\epsscale{1.15}
\plotone{fig4-1-2.pdf}
\caption{Continued.
\label{fig:z48-704-all-2}}
\end{figure}

\begin{figure}[h]
\figurenum{16}
\epsscale{1.15}
\plotone{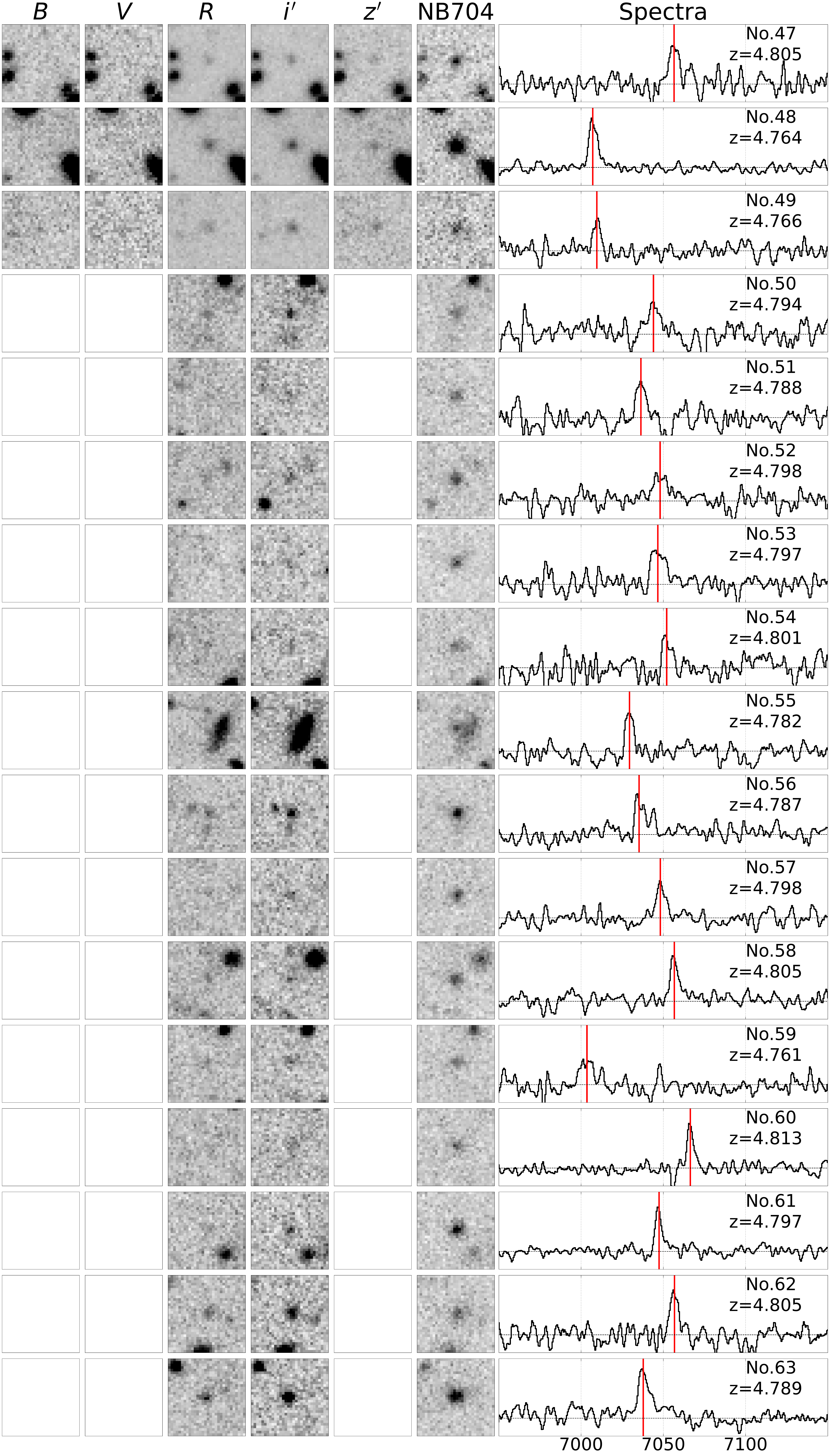}
\caption{Continued.
\label{fig:z48-704-all-3}}
\end{figure}

\begin{figure}[h]
\epsscale{1.15}
\plotone{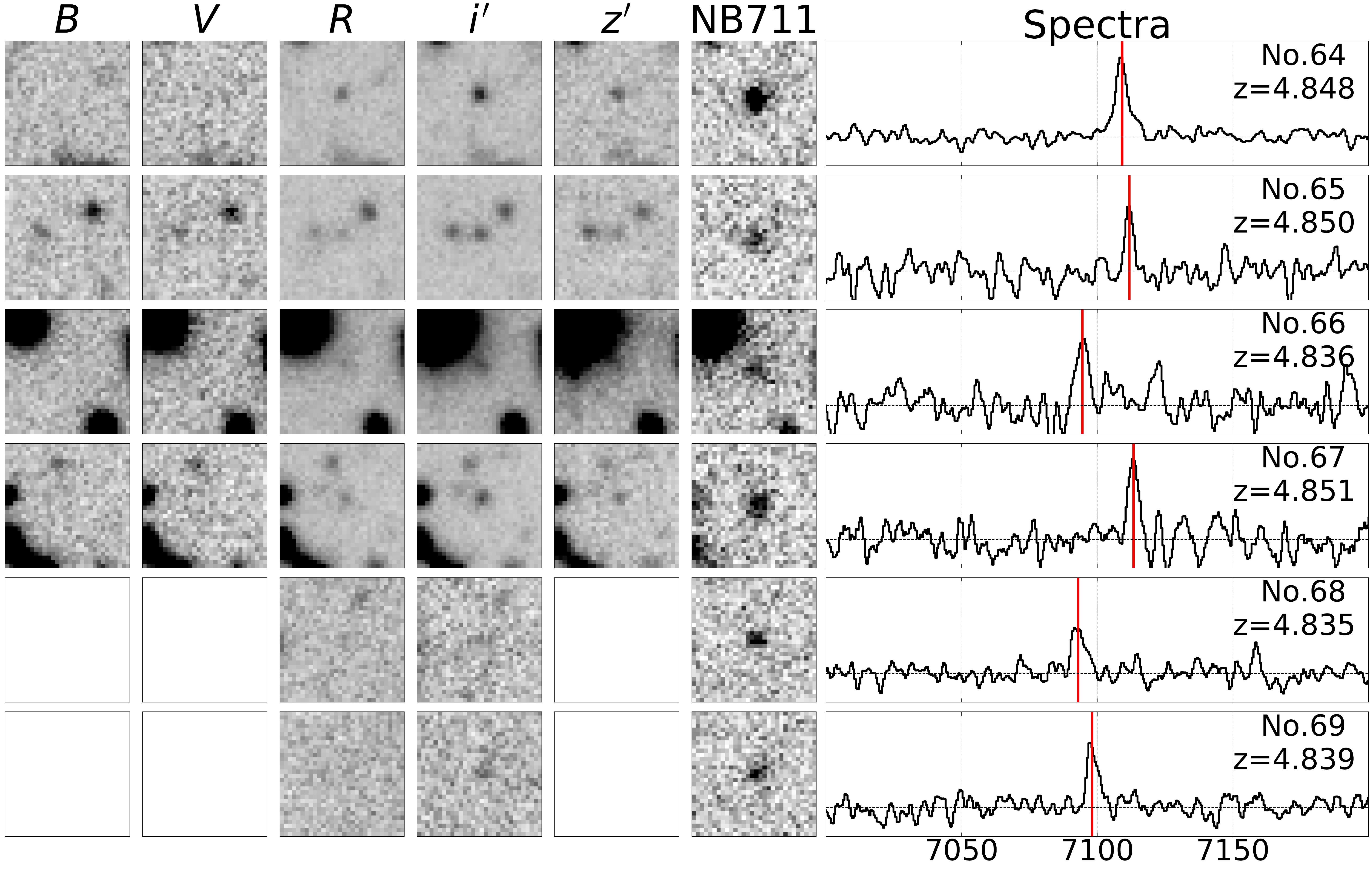}
\caption{Same as Figure \ref{fig:z48-704}, but for all $z\sim4.8$ LAEs selected with narrow band NB711. Objects without $B$, $V$, and $z'$ images are in the SDFn.
\label{fig:z48-711-all-1}}
\end{figure}

\end{document}